\documentclass[draftcls,onecolumn,12pt]{IEEEtran}
\usepackage{bm}
\usepackage{epsfig}
\usepackage{stfloats}
\usepackage{graphicx}
\usepackage{subfigure}
\usepackage{epstopdf}
\usepackage{multirow}
\usepackage[sort, compress]{cite}
\usepackage{epsfig,graphics,subfigure}
\usepackage[cmex10]{amsmath}

\usepackage{array}
\usepackage{tabulary}
\usepackage{booktabs}
\usepackage{amsfonts}
\usepackage{algorithmic}
\usepackage{algorithm}
\usepackage{array}
\usepackage{tabulary}
\usepackage{booktabs}
\usepackage{amsfonts}

\usepackage{algorithmic}
\usepackage{algorithm}

\usepackage{epstopdf}
\def\BibTeX{{\rm B\kern-.05em{\sc i\kern-.025em b}\kern-.08em
    T\kern-.1667em\lower.7ex\hbox{E}\kern-.125emX}}

\usepackage{amssymb}

\usepackage{textcomp}
\usepackage{xcolor}

\begin{document}

\bibliographystyle{IEEEtran}
\title{Covert Surveillance via Proactive Eavesdropping Under Channel Uncertainty}
\author{\IEEEauthorblockN{ Zihao Cheng, Jiangbo Si, {\emph{Member, IEEE}}, Zan Li, {\emph{Senior Member, IEEE}}, Danyang Wang, and Naofal Al-Dhahir, \emph{Fellow, IEEE}}
\thanks{}
\thanks{Jiangbo Si, Zihao Cheng, Danyang Wang, and Zan Li are with the Integrated Service Networks Lab
of Xidian University, Xi'an, 710071, China. Jiangbo Si is also a visiting scholar at the University of Texas at Dallas. (e-mail: jbsi@xidian.edu.cn).}
\thanks{N. Al-Dhahir is with the Department of Electrical and Computer Engineering, The University of Texas at Dallas, Richardson,
TX 75080 USA.(e-mail: aldhahir@utdallas.edu).}}
 \maketitle
\begin{abstract}
Surveillance performance is studied for a wireless eavesdropping system, where a full-duplex legitimate monitor eavesdrops a suspicious link efficiently with the artificial noise (AN) assistance. Different from the existing work in the literature, the suspicious receiver in this paper is assumed to be capable of detecting the presence of AN. Once such receiver detects the AN, the suspicious user will stop transmission, which is harmful for the surveillance performance. Hence, to improve the surveillance performance, AN should be transmitted covertly with a low detection probability by the suspicious receiver. Under these assumptions, an optimization problem is formulated to maximize the eavesdropping non-outage probability under a covert constraint. Based on the detection ability at the suspicious receiver, a novel scheme is proposed to solve the optimization problem by  iterative search. Moreover, we investigate the impact of both the suspicious link uncertainty and the jamming link uncertainty on the covert surveillance performance. Simulations are performed to verify the analyses. We show that the suspicious link uncertainty benefits the surveillance performance, while the jamming link uncertainty can degrade the surveillance performance.

\end{abstract}
\IEEEpeerreviewmaketitle

\section{Introduction}

Physical layer security has been proposed as a promising solution to achieve perfect wireless secrecy against malicious eavesdropping attacks (see e.g. \cite{6327665,7018202,5288936} and references therein). These works often assume that communication users are legitimate and view the information eavesdropping as malicious attacks. However, eavesdropping can also be legitimate behavior. For example, when illegitimate users can potentially establish wireless communication links and pose significant threats to national security, authorized parties must legitimately monitor suspicious communication links for preventing crimes or terrorism attacks.

Many papers have been published on surveillance systems in wireless networks, such as \cite{7321779,7924390,7880684,7544447,7981321,7948745,8254969,8586927}. When the monitor is far away from a suspicious transmitter, it is challenging to eavesdrop the suspicious link. To overcome this problem, a novel approach, namely proactive eavesdropping via artificial noise (AN), was proposed as an efficient way to improve the surveillance performance. For example, a proactive full-duplex monitor purposely injects the AN to decrease the achievable data rate at the suspicious receiver and facilitate  simultaneous eavesdropping \cite{7321779}, where the self-interference between the jamming antennas and the eavesdropping antennas is neglected at the monitor. By contrast, considering the severe self-interference between the jamming antennas and the eavesdropping antennas \cite{7544447,7924390,7880684}, the AN power at the monitor is optimized to maximize the average eavesdropping rate. However, these works do not consider the AN detection ability at the suspicious users. In practice, the suspicious users can be intelligent and have the ability to detect the AN. Thus, if the AN power is large, the suspicious users can detect the AN and stop their transmissions. As a result, the monitor cannot catch the illegitimate information at all.
Hence, the AN should be injected covertly without being detected by the suspicious users.

Covert transmission is an emerging and cutting-edge communication security technique, which guarantees a negligible detection probability at a warden \cite{7355562}. In this regard, the authors in \cite{6584948} established the fundamental limits of covert communication in the form of a square root law, which states that Alice could transmit no more than $O\left( {\sqrt n } \right)$ bits rate  to Bob in $n$ channel uses covertly and reliably. In particular, when $n$ grows to infinity, the rate approaches zero, i.e., ${\lim _{n \to \infty }}\frac{{O\left( {\sqrt n } \right)}}{n} = 0$.
Extensions of this work considered different channels, such as discrete memoryless channels (DMCs) \cite{7447769}, binary symmetric channels (BSCs) \cite{6620765} and multiple access channels (MACs) \cite{7541695}. Fortunately, such a pessimistic conclusion does not hold if the warden has uncertainty about the statistics of the background noise \cite{7084182}.
Meanwhile, the authors of \cite{7805182} investigated the impact of noise uncertainty on detection error probability (DEP) at the warden.
Their results show that noise uncertainty is beneficial to covert transmission.
In order to further enhance the covert performance, cooperative jamming was also employed to increase the noise uncertainty \cite{7964713,8422940,8412160,8654724}. In \cite{7964713}, the authors  employed uninformed jamming to achieve a positive covert transmission rate.
The jamming uncertainty was investigated in \cite{8422940,8412160}, where the locations of the jamming transmitters follow a stationary Poisson point process (PPP).
The performance analysis for covert transmission was also extended to multi-antenna systems in \cite{8849607,8654724}. In addition, a full-duplex receiver injected AN and received messages simultaneously in \cite{8422941,8519751}, where the AN power was designed to achieve the desired level of covertness. Furthermore, relay technology was deployed to enhance covert transmission performance in \cite{8355734,8254008,8736032}.

%

In addition, the channel uncertainty at the warden can also affect the covert transmission performance. Most recently, the impact of channel uncertainty on covert transmission performance was studied in \cite{8108525,8471218,8761935}. With regard to channel uncertainty, the channel state information (CSI) can be separated into the known part and the uncertain part at the receiver. The warden can determine the optimal detection threshold to cancel the known part.
In \cite{8108525}, the authors derived a closed-form expression for the optimal detection threshold and quantified the achievable rate. It is shown in \cite{8108525} that channel uncertainty could help hide the communication to a covert user. The channel uncertainty was also exploited in \cite{8471218} to achieve covert communication in relay networks, where the introduced channel uncertainty confused the warden and limited the ultimate detection performance. In \cite{8761935}, the simulation results show the channel uncertainty has a greater effect on covert transmission rate when the noise uncertainty is larger.

Overall, in order to eavesdrop a suspicious communication link efficiently, the legitimate monitor should inject AN covertly, and control the AN transmission power carefully. However, to the best of the authors' knowledge, intelligent suspicious users, which are capable of detecting a jamming signal and can be common in future wireless networks, have not been well-understood in the literature. Motivated by the above considerations, in this paper, we first study the detection ability at the suspicious users. Then, a novel scheme is proposed to maximize the eavesdropping non-outage probability by optimizing the injected AN power. Moreover, we also investigate the impact of channel uncertainty on covert surveillance performance. The main contributions of this paper are summarized as follows:
\begin{itemize}
\item[1.]
We consider a wireless surveillance system,  where a full-duplex legitimate monitor tries to eavesdrop on a suspicious link with AN assistance. Different from the previous works, such as \cite{7880684,7924390}, the suspicious receiver in this paper has the ability of detecting the AN signal, which is a reasonable assumption in advanced receivers, such as cognitive and military-grade receivers. Specifically, if the suspicious receiver detects the AN from the legitimate monitor, it informs the suspicious transmitter to stop. As a result, the legitimate monitor can overhear nothing. Hence, the legitimate monitor should inject AN covertly. For this scenario, the performance of the surveillance system under a covert constraint is investigated in this paper for the first time, to the best of our knowledge.

\item[2.]
For the surveillance system, an optimization problem is formulated to maximize the eavesdropping non-outage probability while the minimum DEP at the suspicious receiver is less than a given threshold. Closed-form expressions for the optimal detection threshold and the minimum DEP are derived under quasi-static channel fading. Then, based on these expressions, the covert constraint is transformed to an AN power constraint at the legitimate monitor. Finally, we propose an algorithm to achieve the maximum eavesdropping non-outage probability. Numerical results show that the proactive eavesdropping scheme under a covert constraint substantially outperforms the two benchmark schemes in \cite{7321779}.

\item[3.]
Considering that channel uncertainty is inevitable in practice, we analyze the suspicious receiver's AN detection ability under channel uncertainties first in the suspicious link and then in the jamming link. Interestingly, we find that the AN detection ability decreases with the suspicious link channel uncertainty, while it has a non-monotonic relationship with the jamming link channel uncertainty. For the extreme case of perfect suspicious link channel knowledge, the legitimate monitor cannot inject AN to assist its eavesdropping. By contrast, when there exists channel uncertainty for the suspicious link,  even if the suspicious receiver has perfect knowledge of the jamming link, the legitimate monitor can still inject AN without being detected.

\item[4.]We investigate the impact of channel uncertainty on the non-outage probability for the proactive eavesdropping scheme. Numerical results reveal that  the suspicious link channel uncertainty is beneficial to covert surveillance performance. By contrast, the jamming link channel uncertainty affects the surveillance system performance only when the AN power introduces a higher level of interference power at the suspicious receiver than the self-interference power at the legitimate full-duplex monitor. Moreover, since the jamming link channel uncertainty has a non-monotonic effect on the AN detection ability, the legitimate monitor can actively expose the channel knowledge of the jamming link to weaken the suspicious user's detection ability and improve the non-outage probability for the proactive eavesdropping.

\end{itemize}

The rest of this paper is organized as follows.
Section II describes the covert surveillance system model and assumptions. Section III investigates the problem of covert AN transmission.
The surveillance performance under a covert AN constraint is analyzed in Section IV. Special cases are discussed in Section V.
Numerical results are analyzed in Section VI.
Finally, we conclude the paper in Section VII.

\emph{Notations}- $f_\upsilon\left(  \cdot  \right)$ is the probability density function (PDF) of  random variable (RV) $\upsilon$; $F_\upsilon\left(  \cdot  \right)$ is the cumulative distribution
function (CDF) of RV $\upsilon$; $\mathbb{E}{(\cdot)}$ is the expected value of a RV; $\Gamma(\mu, \sigma^2)$ denotes the gamma PDF with
shape $\mu$ and scale $\sigma^2$; $\exp(\sigma^2)$ denotes the exponential PDF with mean $\sigma^2$;$Ei\left( \cdot \right)$ denotes the exponential integral function. In addition, $\chi ^2 \left( {{\nu}} \right)$ denotes the central chi-square PDF with $\nu$ degrees of freedom;
$x \sim {\cal{CN}}\left( {\Lambda ,\Delta } \right)$ denotes the circular symmetric complex Gaussian random vector with mean vector $\Lambda$ and covariance matrix $\Delta$. In addition, the symbol notations are given in Table $1$.

\begin{table}
\caption{Summary of Notations}
\centering
\tiny
\begin{tabular}{|c|c|c|}
\hline
Notation & Description \\
\hline
$h_{i,j}$ & The instantaneous CSI between i and j, $i \in \{{Alice, Monitor}\}$, $j \in \{{Bob, Monitor}\}$ \\
\hline
$\widehat h_{i,j}$ & The perfect known part of $h_{i,j}$ \\
\hline
$\widetilde h_{i,j}$ & The uncertain part of $h_{i,j}$ \\
\hline
$\rho_{i,j}$ & The correlation coefficient between $h_{i,j}$ and $\widetilde h_{i,j}$ \\
\hline
$\sigma^2_{i,j}$ & The covariance of $h_{i,j}$ \\
\hline
$\sigma^2_{j}$ & The noise covariance at node j \\
\hline
$\Gamma$ & The detection threshold at Bob \\
\hline
$\Gamma^*$ & The optimum detection threshold at Bob \\
\hline
$\xi$ & The detection error probability at Bob \\
\hline
$\xi^*$ & The minimum detection probability at Bob \\
\hline
$\bar\xi^*$ & The average minimum detection probability at Bob \\
\hline
$P_{a}$ & The transmission power at Alice \\
\hline
${P_J^{\max }}$ & The maximum transmission power at Monitor \\
\hline
${P_J}$ & The transmission power at Monitor \\
\hline
${P^*_J}$ & The Optimum transmission power at Monitor \\
\hline
${P_J^{covert}}$ & The transmission power at Monitor satisfying covert constraint\\
\hline
${R_M}$ & The achievable rate at Monitor \\
\hline
${R_B}$ & The achievable rate at Bob \\
\hline
${\bf{X}}$ & An indicator function to denote the event of
successful eavesdropping at Monitor \\
\hline
$\delta$ & A  predetermined
threshold for the covert transmission requirement \\
\hline
$\eta$ & The coefficient for the self-interference passive suppression \\
\hline
\end{tabular}
\end{table}

\section{System Model}

\begin{figure}[h]
\centering
\includegraphics[width=4.2in]{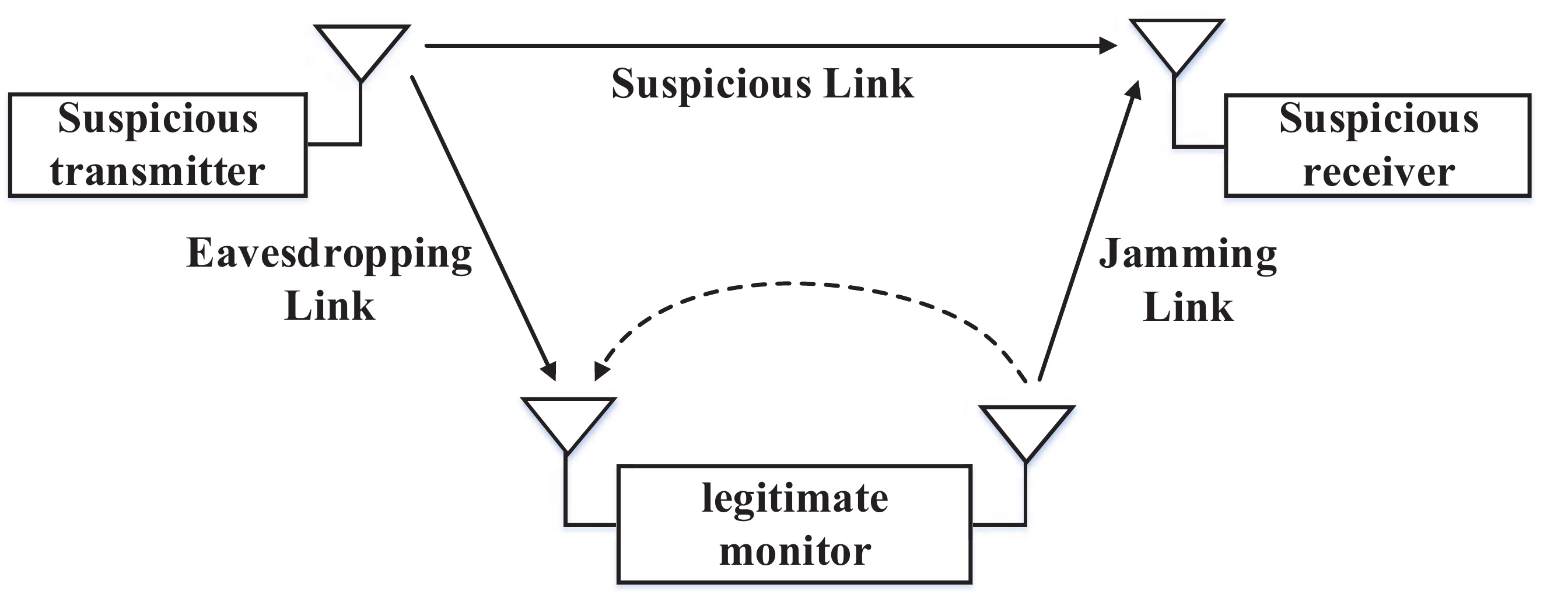}
\caption{Covert surveillance system model} \label{fig.1}
\end{figure}
We consider a covert surveillance system as shown in Fig.1, where a legitimate monitor (Monitor) aims to eavesdrop a suspicious communication link between the suspicious transmitter (Alice) and suspicious receiver (Bob). It is assumed that Alice and Bob are each equipped with a single antenna, while  Monitor is equipped with two antennas and operates in a full-duplex mode. Specifically, Monitor injects AN towards Bob and eavesdrops the signal from Alice simultaneously. In addition, the suspicious receiver Bob not only receives the information signal from Alice, but also detects the jamming signals from Monitor.

Assume that all channels are subject to quasi-static channel fading, and the channel coefficients remain unchanged during each transmission block but vary independently for different fading blocks. The CSIs (i.e., channel coefficients) of the Suspicious Link (Alice $\rightarrow$ Bob), Eavesdropping Link (Alice $\rightarrow$ Monitor), Jamming Link (Monitor $\rightarrow$ Bob) and Self-Interference Link are, respectively, denoted as $h_{A,B}$, $h_{A,M}$, $h_{M,B}$ and $h_{M,M}$, which are complex Gaussian RVs with zero mean and variances $\sigma^2_{A,B}/2$, $\sigma^2_{A,M}/2$, $\sigma^2_{M,B}/2$, and $\sigma^2_{M,M}/2$, respectively.
The transmission power at Alice is $P_{a}$, which is fixed and publicly known by Bob and Monitor. The AN power at Monitor is $P_J$  satisfying $0 \leq P_J \leq P_J^{\max }$ and ${P_J^{\max }}$ denotes the maximum allowed AN power at Monitor.
Then, when Alice transmits the messages, the received signals at Bob and Monitor are given by
\begin{align}\label{yb}\
{y_B}\left[ i \right] = \sqrt {{P_a}} {h_{A,B}}{X_a}\left[ i \right] + \sqrt {{P_J}} {h_{M,B}}{J_a}\left[ i \right] + {n_b}\left[ i \right]
\end{align}
and
\begin{align}\label{ym}\
{y_M}\left[ i \right] = \sqrt {{P_a}} {h_{A,M}}{X_a}\left[ i \right] + \sqrt {{P_J}} \sqrt \eta  {h_{M,M}}{J_a}[i] + {n_m}\left[ i \right],
\end{align}
 where $X_a$ is the transmitted signal from Alice satisfying $E\left[ {{X_a}{{\left[ i \right]}^2}} \right] = 1$; $i = 1,2,...,n$ is the index of each channel use and $J_a$ is the AN from Monitor satisfying $E\left[ {{J_a}{{\left[ i \right]}^2}} \right] = 1$. As in \cite{7924390}, the residual self-interference channel is modeled by $\sqrt \eta  {h_{M,M}}$, where $\eta \left( {0 \le \eta \le 1} \right)$  parameterizes the effect of passive self-interference suppression. Finally, ${n_b}\left[ i \right]$ is the additive white Gaussian noise (AWGN) at Bob with variance $\sigma^2_{b}$, i.e., ${n_b}\left[ i \right] \sim {\cal{CN}}\left( {0,\sigma _b^2} \right)$; ${n_m}\left[ i \right]$ is the AWGN at Monitor with variance $\sigma^2_{m}$, i.e. ${n_m}\left[ i \right] \sim {\cal{CN}}\left( {0,\sigma _m^2} \right)$.

\subsection{Channel Knowledge}
Since the channel estimation
problem is generally not error-free, we assume that $h_{A,B}$ can be separated into the known part ${\widehat h_{A,B}}$ and the uncertain part ${\widetilde h_{A,B}}$ as in \cite{8108525}, and it is given by
\begin{align}\label{p_out}\
{h_{A,B}} = {\widehat h_{A,B} + \widetilde h_{A,B}},
\end{align}
where $\widehat h_{A,B}$ and $ \widetilde h_{A,B}$ are independent complex Gaussian random variables with mean zero and variances $\left( {1 - {\rho _{a,b}}} \right)\sigma _{A,B}^2$ and $\rho_{a,b} \sigma _{A,B}^2$, respectively. $\rho_{a,b} \in \left[ {0,1} \right]$ denotes the correlation coefficient between  $\widetilde h_{A,B}$ and ${h_{A,B}}$. Namely, when $\rho_{a,b}=0$, $h_{A,B}$ is perfectly known by Bob.

Usually, we assume that Bob can only know the statistical CSI of the Jamming Link $h_{M,B}$, and rely on the average measure of his performance to detect the AN from Monitor \cite{8108525} \cite{8471218}. Nevertheless, since Monitor injects AN, the instantaneous CSI of the jamming link can be tracked by Bob \cite{4529282,5485016,6623091}. Due to the imperfect channel estimation, we specifically investigate the channel uncertainty of $h_{M,B}$. In particular, $h_{M,B}$ at Bob can also be separated into the known part ${\widehat h_{M,B}}$ and the uncertain part ${\widetilde h_{M,B}}$, and it is given by
\begin{align}\label{p_out}\
{h_{M,B}} = {\widehat h_{M,B} + \widetilde h_{M,B}},
\end{align}
where $\widehat h_{M,B}$ and $ \widetilde h_{M,B}$ are independent complex Gaussian random variables with mean zero and variances $\left( {1 - {\rho _{m,b}}} \right)\sigma _{M,B}^2$ and $\rho_{m,b} \sigma _{M,B}^2$, respectively. $\rho_{m,b} \in \left[ {0,1} \right]$ denotes the correlation coefficient between $ \widetilde h_{M,B}$ and ${h_{M,B}}$.
\subsection{Hypothesis Testing}
In order to detect the AN from Monitor, Bob faces a binary hypothesis testing problem. We consider two events: $H_0$ and $H_1$, where $H_0$ denotes the null hypothesis when Monitor does not transmit AN, and $H_1$ denotes the alternative hypothesis when Monitor transmits AN to Bob. For the two cases, the received signal at Bob is given by
\begin{align}\label{ybh0h1}\
{y_B}\left[ i \right] = \left\{ {\begin{array}{*{20}{c}}
{\sqrt {{P_a}} {h_{A,B}}{X_a}\left[ i \right] + {n_b}\left[ i \right]}&{{H_0}}\\
{\sqrt {{P_a}} {h_{A,B}}{X_a}\left[ i \right] + \sqrt {{P_J}} {h_{M,B}}{J_a}\left[ i \right] + {n_b}\left[ i \right]}&{{H_1}}.
\end{array}} \right.
\end{align}
By application of the Neyman-Pearson criterion \cite{8108525}, the decision rule for Bob is given by
\begin{align}\label{tn}\
T\left( n \right) = \frac{1}{n}\sum\nolimits_{i = 1}^n {{{\left| {{y_B}\left[ i \right]} \right|}^2}\mathop  \lessgtr \limits_{{D_1}}^{{D_0}} } {\kern 1pt} {\kern 1pt} {\kern 1pt} {\kern 1pt} \Gamma,
\end{align}
where $\Gamma$ is the detection threshold; $D_0$ and $D_1$ are the decisions in favor of $H_0$ and $H_1$, respectively. In this paper, we consider an infinite number of channel uses i.e., $n\rightarrow\infty $. Thus, the average power received at Bob is given by
\begin{align}\label{tnh0h1}\
{T_B} = \left\{ {\begin{array}{*{20}{c}}
{{P_a}{{\left| {{h_{A,B}}} \right|}^2} + \sigma _b^2}&{{H_0}}\\
{{P_a}{{\left| {{h_{A,B}}} \right|}^2} + {P_J}{{\left| {{h_{M,B}}} \right|}^2} + \sigma _b^2}&{{H_1}},
\end{array}} \right.
\end{align}
Since ${\left| {{{\widehat h}_{A,B}} + {{\widetilde h}_{A,B}}} \right|^2}$ can be approximated as  ${\left| {{{\widehat h}_{A,B}}} \right|^2} + {\left| {{{\widetilde h}_{A,B}}} \right|^2}$, and ${\left| {{{\widehat h}_{M,B}} + {{\widetilde h}_{M,B}}} \right|^2}$ can be approximated as ${\left| {{{\widehat h}_{M,B}}} \right|^2} + {\left| {{{\widetilde h}_{M,B}}} \right|^2}$\cite{8761935,8108525}, \eqref{tnh0h1} can be reformulated as
\begin{align}\label{tnh0h2}\
{T_B} = \left\{ {\begin{array}{*{20}{c}}
{{P_a}{{\left| {{{\hat h}_{A,B}}} \right|}^2} + {P_a}{{\left| {{{\widetilde h}_{A,B}}} \right|}^2} + \sigma _b^2}&{{H_0}}\\
{{P_a}{{\left| {{{\hat h}_{A,B}}} \right|}^2} + {P_a}{{\left| {{{\widetilde h}_{A,B}}} \right|}^2} + {P_J}{{\left| {{{\hat h}_{M,B}}} \right|}^2} + {P_J}{{\left| {{{\widetilde h}_{M,B}}} \right|}^2} + \sigma _b^2}&{{H_1}}.
\end{array}} \right.
\end{align}

In this paper, the false alarm probability and the missed detection probability are defined as $P_{FA} \buildrel \Delta \over = \Pr \left( {{D_1}|{H_0}} \right)$ and $P_{MD} \buildrel \Delta \over = \Pr \left( {{D_0}|{H_1}} \right)$.
Since the false alarm and the missed detection events are the two types of errors for Bob's hypothesis test, under the assumption of equal probability for $H_0$ and $H_1$, the performance of Bob's hypothesis test is measured by
\begin{align}\label{gamma_M}\
\xi  = P_{FA} + P_{MD},
\end{align}
where $\xi$ denotes the detection error probability (DEP). $\xi=0$ means that Bob can detect AN without error, and $\xi=1$ means that Bob cannot detect AN from Monitor at all.  From the prosperity of Bob, it will determine the optimal detection threshold $\Gamma^*$  and obtain the minimum DEP $\xi^*$ based on the estimated CSI of the Suspicious link and Jamming link, i.e., ${{P_a}{{\left| {{{\hat h}_{A,{B}}}} \right|}^2}}$ and ${{P_J}{{\left| {{{\hat h}_{M,B}}} \right|}^2}}$. From the perspective of Monitor, it will inject AN under the assumption that Bob has the best AN detection performance, i.e. the worst case for Monitor is addressed.

\section{Injecting AN Covertly}
When Bob detects the interference from Monitor, Alice will stop transmission. This will degrade the surveillance performance. Hence, to improve the surveillance performance and prevent AN from being detected, the monitor should inject AN covertly and control the AN transmission power carefully. In this section, the worst case is considered where we assume that Bob has the best detection ability. The optimal detection threshold ${\Gamma ^ * }$ and minimum DEP ${{\xi ^*}} $ at Bob are firstly derived under the channel uncertainties of $h_{A,B}$ and $h_{M,B}$. Then, the maximum permitted transmission AN power at the monitor is determined under the covert constraint. For the optimal detection threshold ${\Gamma ^ * }$ and minimum DEP ${{\xi ^*}}$ at Bob, we have the following lemma.

{\emph{\textbf{Lemma 1:}}} ${\Gamma ^ * }$ and ${{\xi ^*}}$ are, respectively, given by
\begin{align}\label{optimal_menxian}\
{\Gamma ^*} = {X_1} + {X_2} + {k_1} + \sigma _b^2
\end{align}
and
\begin{align}\label{optimal_DEP}\
{\xi ^*} &= 1 + \exp \left( { - \frac{{{X_2} + {k_1}}}{{{\rho _{a,b}}{P_a}{\sigma ^2}_{A,B}}}} \right) \nonumber \\
 & \quad - \frac{{{\rho _{a,b}}{P_a}{\sigma ^2}_{A,B}\exp \left( { - \frac{{{k_1}}}{{{\rho _{a,b}}{P_a}{\sigma ^2}_{A,B}}}} \right) - {\rho _{m,b}}{P_J}{\sigma ^2}_{M,B}\exp \left( { - \frac{{{k_1}}}{{{\rho _{m,b}}{P_J}{\sigma ^2}_{M,B}}}} \right)}}{{{\rho _{a,b}}{P_a}{\sigma ^2}_{A,B} - {\rho _{m,b}}{P_J}{\sigma ^2}_{M,B}}},
\end{align}
where ${X_1} = {P_a}{\left| {{{\hat h}_{A,B}}} \right|^2}$, ${X_2} = {P_J}{\left| {{{\hat h}_{M,B}}} \right|^2}$, and \\
${k_1} = \frac{{{\rho _{m,b}}{\rho _{a,b}}{P_J}{P_a}{\sigma ^2}_{M,B}{\sigma ^2}_{A,B}}}{{{\rho _{m,b}}{P_J}{\sigma ^2}_{M,B} - {\rho _{a,b}}{P_a}{\sigma ^2}_{A,B}}}\ln \left( {1 - \left( {1 - \frac{{{\rho _{m,b}}{P_J}{\sigma ^2}_{M,B}}}{{{\rho _{a,b}}{P_a}{\sigma ^2}_{A,B}}}} \right)\exp \left( { - \frac{{{X_2}}}{{{\rho _{a,b}}{P_a}{\sigma ^2}_{A,B}}}} \right)} \right)$.

{\emph{Proof:}}
According to \eqref{tnh0h2}, the average power received at Bob is
 \begin{align}\label{average_power}\
{T_B} = \left\{ {\begin{array}{*{20}{c}}
{X_1+ {P_a}{{\left| {{{\widetilde h}_{A,B}}} \right|}^2} + \sigma _b^2}&{{H_0}}\\
{X_1 + {P_a}{{\left| {{{\widetilde h}_{A,B}}} \right|}^2} + X_2 + {P_J}{{\left| {{{\widetilde h}_{M,B}}} \right|}^2} + \sigma _b^2}&{{H_1},}
\end{array}} \right.
\end{align}
where $X_1$ and $X_2$  are perfectly known by Bob. As in \cite{8108525,8471218}, the probabilities of false alarm and missed detection are, respectively, given by
 \begin{align}\label{PFA}\
{P_{FA}} &= P\left( {{P_a}{{\left| {{{\widetilde h}_{A,B}}} \right|}^2} > \Gamma  - \sigma _b^2 - {X_1}} \right)\nonumber \\
& = \left\{ {\begin{array}{*{20}{c}}
1&{\Gamma  \le {X_1} + \sigma _b^2}\\
{\exp \left(-\frac{ {\Gamma  - \sigma _b^2 - {X_1}}}{{{\rho _{a,b}}{P_a}{\sigma ^2}_{A,B}}}  \right)}&{\Gamma  > {X_1} + \sigma _b^2}
\end{array}} \right.
\end{align}
and
 \begin{align}\label{PMD}\
{P_{MD}} &= P\left( {{P_a}{{\left| {{{\widetilde h}_{A,B}}} \right|}^2} + {P_J}{{\left| {{{\widetilde h}_{M,B}}} \right|}^2} < \Gamma  - \sigma _b^2 - {X_1} - {X_2}} \right)\nonumber\\
 &= \left\{ {\begin{array}{*{20}{c}}
0&{\Gamma  \le {X_1} + {X_2} + \sigma _b^2}\\
{1 - \frac{{{\rho _{a,b}}{P_a}\sigma _{A,B}^2\exp \left( { - \frac{{\Gamma  - \sigma _b^2 - {X_1} - {X_2}}}{{{\rho _{a,b}}{P_a}\sigma _{A,B}^2}}} \right) - {\rho _{m,b}}{P_J}\sigma _{M,B}^2\exp \left( { - \frac{{\Gamma  - \sigma _b^2 - {X_1} - {X_2}}}{{{\rho _{m,b}}{P_J}\sigma _{M,B}^2}}} \right)}}{{{\rho _{a,b}}{P_a}\sigma _{A,B}^2 - {\rho _{m,b}}{P_J}\sigma _{M,B}^2}}}&{\Gamma  > {X_1} + {X_2} + \sigma _b^2.}
\end{array}} \right.
\end{align}

Since $\xi  = {P_{FA}} + {P_{MD}}$, we can obtain the DEP as
 \begin{align}\label{PDEP}\
\xi  = \left\{ {\begin{array}{*{20}{c}}
1&{\Gamma  < {X_1} + \sigma _b^2}\\
{{P_{FA}}}&{{X_1} + \sigma _b^2 \le \Gamma  < {X_1} + {X_2} + \sigma _b^2}\\
{{P_{FA}} + {P_{MD}}}&{{X_1} + {X_2} + \sigma _b^2 \le \Gamma }.
\end{array}} \right.
\end{align}
Clearly, when ${X_1} + \sigma _b^2 \le \Gamma  < {X_1} + {X_2} + \sigma _b^2$, $\xi$ decreases with $\Gamma$. On the other hand, we find that $\xi$ decreases with $\Gamma$ when ${X_1} + {X_2} + \sigma _b^2 \leq  \Gamma  < {X_1} + {X_2} + \sigma _b^2 + {k_1}$, and $\xi$ increases with $\Gamma$ when $\Gamma  > {X_1} + {X_2} + \sigma _b^2 + {k_1}$. Hence, when ${\Gamma ^*} = {X_1} + {X_2} + \sigma _b^2 + k_1$, $\xi $ can achieve the minimum value. The proof of this lemma is completed.

Since Monitor has no knowledge about $X_1$ and $ X_2$,  it has to rely on the average measure of Bob's detection performance to assess the possible covertness as in \cite{8108525}. Note that the PDF of $X_2$ is given by
\begin{align}\label{pdf}\
{f_{{X_2}}}\left( x \right) = \frac{1}{{\left( {1 - {\rho _{m,b}}} \right){P_J}\sigma _{M,B}^2}}\exp \left( { - \frac{x}{{\left( {1 - {\rho _{m,b}}} \right){P_J}\sigma _{M,B}^2}}} \right).
\end{align}
Hence, the average minimum DEP $\overline {{\xi ^*}} $ is calculated as
\begin{align}\label{DEP_average_minmum}\
\overline {{\xi ^*}}  &= \int_0^\infty  {{\xi ^*}{f_{{X_2}}}\left( x \right)} dx\nonumber\\
 &= 1 - {\left( {1 - \frac{{{\rho _{m,b}}{P_J}\sigma _{M,B}^2}}{{{\rho _{a,b}}{P_a}\sigma _{A,B}^2}}} \right)^{ - \frac{{{\rho _{a,b}}{P_a}\sigma _{A,B}^2}}{{\left( {1 - {\rho _{m,b}}} \right){P_J}\sigma _{M,B}^2}}}}\frac{{{\rho _{a,b}}{P_a}\sigma _{A,B}^2}}{{\left( {1 - {\rho _{m,b}}} \right){P_J}\sigma _{M,B}^2}}\nonumber\\
 &\times \int_{\frac{{{\rho _{m,b}}{P_J}\sigma _{M,B}^2}}{{{\rho _{a,b}}{P_a}\sigma _{A,B}^2}}}^1 {{x^{{{\left( {1 - \frac{{{\rho _{m,b}}{P_J}\sigma _{M,B}^2}}{{{\rho _{a,b}}{P_a}\sigma _{A,B}^2}}} \right)}^{ - 1}}}}{{\left( {1 - x} \right)}^{\frac{{{\rho _{a,b}}{P_a}\sigma _{A,B}^2}}{{\left( {1 - {\rho _{m,b}}} \right){P_J}\sigma _{M,B}^2}} - 1}}} dx.
\end{align}
$\overline{\xi^*}$ decreases with $\bar{P}_J$ which is proved in Appendix A. To provide more insight, Fig. 2 shows $\overline {{\xi ^*}}$ for a given $\rho_{m,b}$ versus the normalized power $\bar{P}_J={P}_J/\sigma^2_b$. The exactness of \eqref{DEP_average_minmum} is verified in Fig. 2. Furthermore, we get the relationship between $\overline{\xi^*}$ and $\rho_{a,b}$, as well as with $\rho_{m,b}$ in the following remark.

{\emph{Remark 1:}}
 $\overline{\xi^*}$ increases with $\rho_{a,b}$. By contrast, $\rho_{m,b}$ has a non-monotonic effect on $\overline{\xi^*}$. This indicates that the Suspicious Link channel uncertainty degrades Bob's AN detection ability, while the Jamming Link channel uncertainty can degrade or benefit Bob's detection ability. Specifically, when the jamming power is low, the jamming link uncertainty is beneficial to the covert transmission. But when the jamming power is large, the jamming link uncertainty is detrimental to the covert transmission because the optimal detection threshold is affected. Moreover, the channel uncertainty can affect the AN power at Monitor by affecting Bob's AN detection ability, so as to affect the surveillance performance.

 In practice, there are several factors affecting the uncertainty of the Suspicious link. For example, high dynamic channel in time, frequency, and space domain, and receiver capability. It is an effective way to improve the covert transmission performance by frequently changing the electronic environment at Bob. As a result, Bob is difficult to get the perfect knowledge of the Suspicious link. In addition, continuously changing the transmission power is also an attractive method to increase the jamming link uncertainty.
\begin{figure}[ht]
\centering
\includegraphics[width=0.6\textwidth]{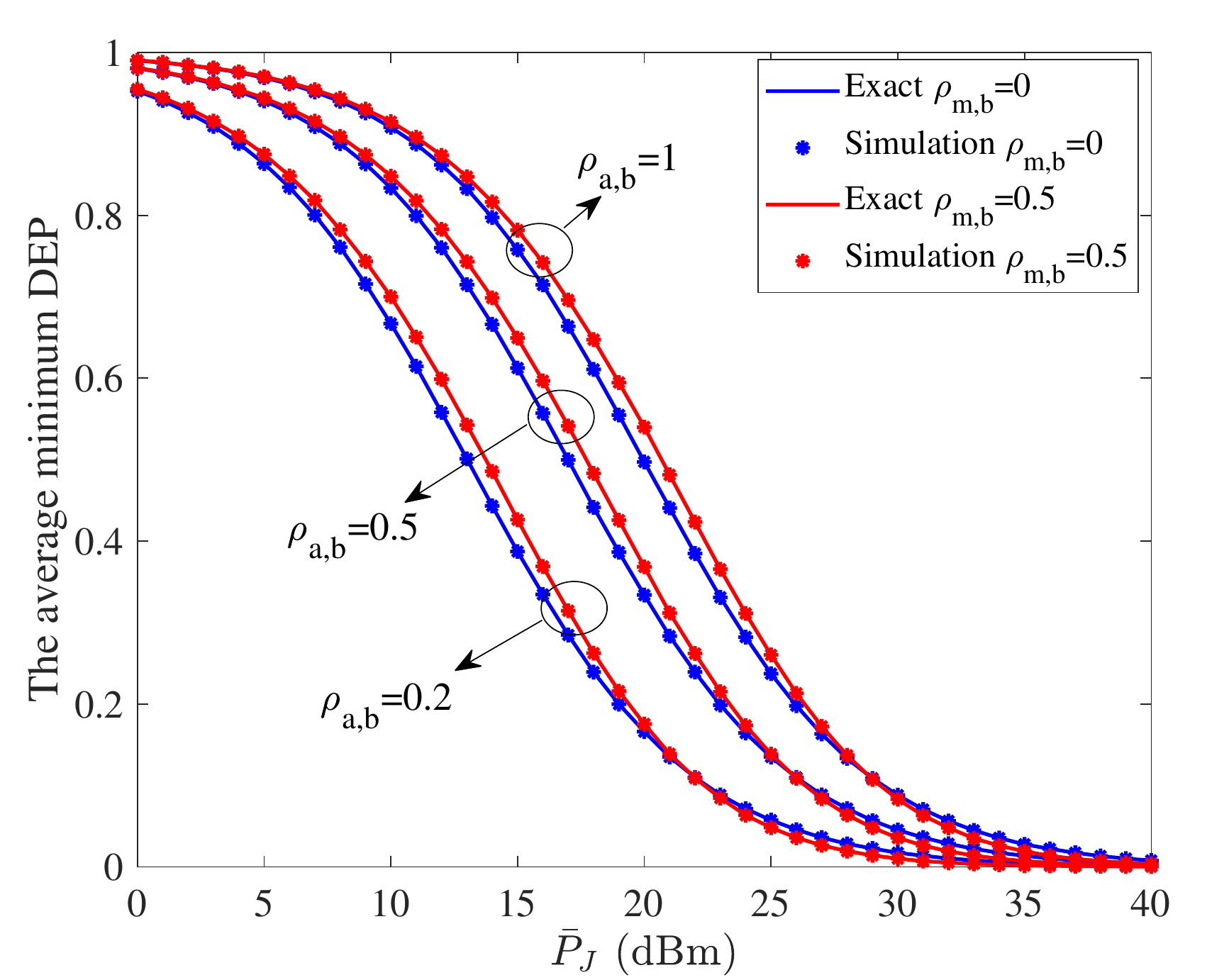}
\caption{The average minimum DEP $\overline {{\xi ^*}} $ against $\rho_{a,b}$ for different $\rho_{m,b}$.} \label{Fig3}
\end{figure}

In this paper, covert transmission is assumed $\overline {{\xi ^*}}\geq 1-\delta$, where $\delta$ denotes a predetermined threshold for the covert transmission requirement. Note that when $\overline {{\xi ^*}}=1-\delta$, we obtain $P_J^{covert} ={\left(\overline {{\xi ^*}}\right) ^{ - 1}}\left( {1-\delta} \right)$, where ${\left(\overline {{\xi ^*}}\right) ^{ - 1}}\left( {1-\delta} \right)$ is the inverse function of $\overline {{\xi ^*}}$.
If the AN power at Monitor is higher than $P_J^{covert}$, the covert constraint cannot be satisfied. Thus, to satisfy the covert constraint, the allowable range of $P_J$ is given by
\begin{align}\label{value_range}\
0 \le {P_J} \le \min \left( {P_J^{\max },P_J^{covert}} \right).
\end{align}
According to \eqref{value_range}, the covert constraint is transformed to an AN  transmission power constraint at Monitor.

\section{Surveillance Performance Under Covert Constraint}
When the AN power injected from Monitor is controlled to satisfy the covert constraint, Monitor can overhear the suspicious link efficiently. There are several  metrics for wireless surveillance performance evaluation, such as the eavesdropping non-outage probability \cite{7321779,7924390}, the average monitor rate \cite{7981321,7948745}, and the successful eavesdropping rate or the effective eavesdropping rate \cite{8254969,8586927}.

\subsection{Eavesdropping Non-outage Probability}
In this paper, for simplicity, we adopt the eavesdropping non-outage probability as a performance metric, which is defined as $E\left[ {\bf{X}} \right]$. $\bf{X}$ is an indicator function to denote the event of successful eavesdropping at Monitor as in \cite{7321779}, and it can be expressed as
\begin{align}\label{Rs=RmRb}\
{\bf{X}} = \left\{ {\begin{array}{*{20}{c}}
1&{if\quad{R_M} \ge {R_B}}\\
0&{otherwise},
\end{array}} \right.
\end{align}
where ${\bf{X}} = 1$ denotes eavesdropping non-outage event that Monitor can reliably decode the information, and ${\bf{X}} = 0$ denotes eavesdropping outage event that Monitor cannot perfectly decode the information. $R_B$ and $R_M$ are, respectively, denoted as the achievable rates at Bob and Monitor, and they are given by
\begin{align}\label{R_B}\
{R_B} = {\log _2}\left( {1 + \frac{{{P_a}{{\left| {{h_{A,B}}} \right|}^2}}}{{{P_J}{{\left| {{h_{M,B}}} \right|}^2} + \sigma _b^2}}} \right)
\end{align}
and
\begin{align}\label{R_M}\
{R_M} = {\log _2}\left( {1 + \frac{{{P_a}{{\left| {{h_{A,M}}} \right|}^2}}}{{\eta {P_J}{{\left| {{h_{M,M}}} \right|}^2} + \sigma _m^2}}} \right).
\end{align}

To successfully eavesdrop the information of the Suspicious Link, Monitor should ensure that its achievable data rate $R_M$ is greater than Bob's rate $R_B$. According to \eqref{Rs=RmRb}, $E\left[ {\bf{X}} \right]$ can be formulated as
\begin{align}\label{ENP}\
{\bf{E}}\left[ \bf{X} \right] &= Prob\left( {\frac{{{P_a}{{\left| {{h_{A,M}}} \right|}^2}}}{{\eta{P_J}{{\left| {{h_{M,M}}} \right|}^2} + \sigma _m^2}} \ge \frac{{{P_a}{{\left| {{{\hat h}_{A,B}}} \right|}^2}}}{{{P_a}{{\left| {{{\widetilde h}_{A,B}}} \right|}^2} + {P_J}{{\left| {{{\widetilde h}_{M,B}} + {{\widehat h}_{M,B}}} \right|}^2} + \sigma _b^2}}} \right)\nonumber\\
 &= Prob\left( {\frac{{{P_a}{{\left| {{h_{A,M}}} \right|}^2}}}{{\eta {P_J}{{\left| {{h_{M,M}}} \right|}^2} + \sigma _m^2}} \ge \frac{{\left( {1 - {\rho _{a,b}}} \right){P_a}{{\left| {{h_{A,B}}} \right|}^2}}}{{{\rho _{a,b}}{P_a}{{\left| {{h_{A,B}}} \right|}^2} + {P_J}{{\left| {{h_{M,B}}} \right|}^2} + \sigma _b^2}}} \right),
\end{align}
From \eqref{ENP}, it is obvious that different values of $\rho_{m,b}$ have no direct effect on $E\left[ {\bf{X}} \right]$. In contrast, $E\left[ {\bf{X}} \right]$ increases with $\rho_{a,b}$.
\subsection{Optimization Problem}
Note that to maximize $E\left[ {\bf{X}} \right]$, the AN transmitted from Monitor should not be detected at Bob. Otherwise, Alice will stop transmission and Monitor can overhear nothing. To this end, the objective for Monitor is to maximize the eavesdropping non-outage probability $E\left[ {\bf{X}} \right]$ while AN is injected covertly. Hence, the optimization problem can be formulated as
\begin{subequations}
\begin{align}\label{problem}\
&\begin{array}{*{20}{c}}
{\mathop {\max }\limits_{\hfill\scriptstyle{P_{{J}}}\hfill\atop} }& E\left[ {\bf{X}} \right]
\end{array}\\
&{s.t.}\quad 0 \leq P_J \leq P_J^{\max },\label{problema}\\
&\quad \quad {\overline{\xi^*}} \ge 1 - \delta \label{problemb},
\end{align}
\end{subequations}
where \eqref{problema} is the AN power constraint. Equation \eqref{problemb} is the covert constraint. Note that the objective function ${\bf{E}}\left[ \bf{X} \right]$ is not concave over the AN power $P_J$. Fortunately, according to \cite[Theorem 1]{7924390}, the optimal AN power $P_J^*$ can be expressed as
\begin{small}
\begin{align}\label{P_J}\
P_J^* = \left\{ {\begin{array}{*{20}{l}}
{\min \left( {P_J^{\max },P_J^{{\mathop{\rm cov}} ert}} \right){\kern 1pt} {\kern 1pt} {\kern 1pt} {\kern 1pt} {\kern 1pt} {\kern 1pt} {\kern 1pt} {\kern 1pt} {\kern 1pt} {\kern 1pt} if\left\{ {\begin{array}{*{20}{l}}
{{\kern 1pt} {\Delta _1} > 0,{\kern 1pt}  {\Delta _2} > 0{\kern 1pt} ,{\kern 1pt} \frac{{{\Delta _2}}}{{{\Delta _1}}} \ge P_J^{\max },and{\kern 1pt}\overline {{\xi ^*}}  \ge 1 - \delta }\\
{{\kern 1pt} {\kern 1pt} {\Delta _1} = 0,{\Delta _2} > 0, {\kern 1pt}\overline {{\xi ^*}}  \ge 1 - \delta }\\
{{\kern 1pt} {\kern 1pt} {\kern 1pt} {\Delta _2} \ge 0\overline {,{\xi ^*}}  \ge 1 - \delta \left\{ {\begin{array}{*{20}{l}}
{{\Delta _2} \ge 0}\\
{{\Delta _2} < 0,\frac{{{\Delta _2}}}{{{\Delta _1}}} < P_J^{\max }, {\kern 1pt} {\kern 1pt} {\kern 1pt} {\kern 1pt} {\kern 1pt} {\kern 1pt} {\Delta _3} \ge {\Delta _4}}
\end{array}} \right.}
\end{array}} \right.}\\
{0{\kern 1pt} {\kern 1pt} {\kern 1pt} {\kern 1pt} {\kern 1pt} {\kern 1pt} {\kern 1pt} {\kern 1pt} {\kern 1pt} {\kern 1pt} {\kern 1pt} {\kern 1pt} {\kern 1pt} {\kern 1pt} if\left\{ {\begin{array}{*{20}{l}}
{{\Delta _1} \ge 0,{\Delta _2} \le 0}\\
\begin{array}{l}
{\Delta _1} < 0,{\Delta _2} < 0\left\{ {\begin{array}{*{20}{l}}
{\frac{{{\Delta _2}}}{{{\Delta _1}}} \ge P_J^{\max }}\\
{\frac{{{\Delta _2}}}{{{\Delta _1}}} < P_J^{\max }, {\kern 1pt} {\Delta _3} < {\Delta _4}}
\end{array}} \right.\\
\overline {{\xi ^*}}  < 1 - \delta
\end{array}
\end{array}} \right.}\\
{\min \left( {\frac{{{\Delta _2}}}{{{\Delta _1}}},P_J^{{\mathop{\rm cov}} ert}} \right){\kern 1pt} {\kern 1pt} {\kern 1pt} {\kern 1pt} {\kern 1pt} {\kern 1pt} {\kern 1pt} {\kern 1pt} {\kern 1pt} {\kern 1pt} {\kern 1pt} {\kern 1pt} {\kern 1pt} {\kern 1pt} {\kern 1pt} {\kern 1pt} {\kern 1pt} if{\Delta _1} > 0,{\kern 1pt} {\kern 1pt} {\Delta _2} > 0,{\kern 1pt} {\kern 1pt} {\kern 1pt} \frac{{{\Delta _2}}}{{{\Delta _1}}} < P_J^{\max }, {\kern 1pt}\overline {{\xi ^*}}  \ge 1 - \delta},
\end{array}} \right.
\end{align}
\end{small}
where ${\Delta _1} = {\left| {{h_{M,B}}} \right|^2} - \frac{{\sqrt \eta \left| {{h_{M,M}}} \right|\left| {{{\hat h}_{A,B}}} \right|\left| {{h_{M,B}}} \right|}}{{\left| {{h_{A,M}}} \right|}}$, ${\Delta _2} = \frac{{\left| {{{\hat h}_{A,B}}} \right|\left| {{h_{M,B}}} \right|}}{{\sqrt \eta  \left| {{h_{M,M}}} \right|\left| {{h_{A,M}}} \right|}}\sigma _m^2 - {P_a}{\left| {{{\widetilde h}_{A,B}}} \right|^2} - \sigma _b^2$, ${\Delta _3} = \frac{{\sigma _m^2{{\left| {{{\hat h}_{A,B}}} \right|}^2}{{\left| {{h_{M,B}}} \right|}^2}}}{{{P_a}{{\left| {{{\widetilde h}_{A,B}}} \right|}^2} + {P_J}{{\left| {{h_{M,B}}} \right|}^2} + \sigma _b^2}}$, and ${\Delta _4} = \frac{{\eta {{\left| {{h_{M,M}}} \right|}^2}{{\left| {{h_{A,M}}} \right|}^2}\left( {{P_a}{{\left| {{{\widetilde h}_{A,B}}} \right|}^2} + \sigma _b^2} \right)}}{{\eta {P_J}{{\left| {{h_{M,M}}} \right|}^2} + \sigma _m^2}}$.
From \eqref{P_J}, we find that $P_J^*$ not only depends on the relationship between the channel gains and noise powers but also on Alice's transmission power, which is different from the existing works \cite{7321779,7544447,7924390,7880684}. This behavior is caused by the channel uncertainty of the Suspicious link. In addition, even if AN introduces a higher level of interference power at Bob than the self-interference power at Monitor, Monitor still needs to consider the covert constraint rather than using full power to confuse Bob. As for the scenario that Monitor can already overhear from the transmitter successfully without AN or when the self-interference is severe, no AN is required and $P_J^*=0$. After we determine the optimal AN power, the following lemma determines the exact eavesdropping non-outage probability $E\left[ {\bf{X}} \right]$.

{\emph{\textbf{Lemma 2:}}} $E\left[ {\bf{X}} \right]$ can be calculated as \eqref{ENP_exact}.

{\emph{Proof:}} See Appendix C. {\text{Lemma 2}} presents a closed-form expression for the eavesdropping non-outage probability.  To provide more insight, Fig. 3 shows the eavesdropping non-outage probability versus $P_J$ without covert constraint. In addition, the exactness of \eqref{ENP_exact} is verified in Fig. 3. We can see that the relationship between $P_J$ and eavesdropping non-outage probability is not monotonous, even if there is no covert constraint. This is due to self-interference effect.

Although it is challenging to obtain the exact expression for the optimal AN power and get the maximum eavesdropping non-outage probability under covert constraint, a simple search method can be sufficient to solve the optimization problem \eqref{problem}. According to \eqref{value_range}, the allowable range of $P_J$ is ensured, thus the maximum $E\left[ {\bf{X}} \right]$ can be obtained by Algorithm 1. In this algorithm, $\Delta {P_J}$ denotes the incremental step for $P_J$, and it can be determined by the required $E\left[ {\bf{X}} \right]$ accuracy.


\begin{figure}[ht]
\centering
\includegraphics[width=0.5\textwidth]{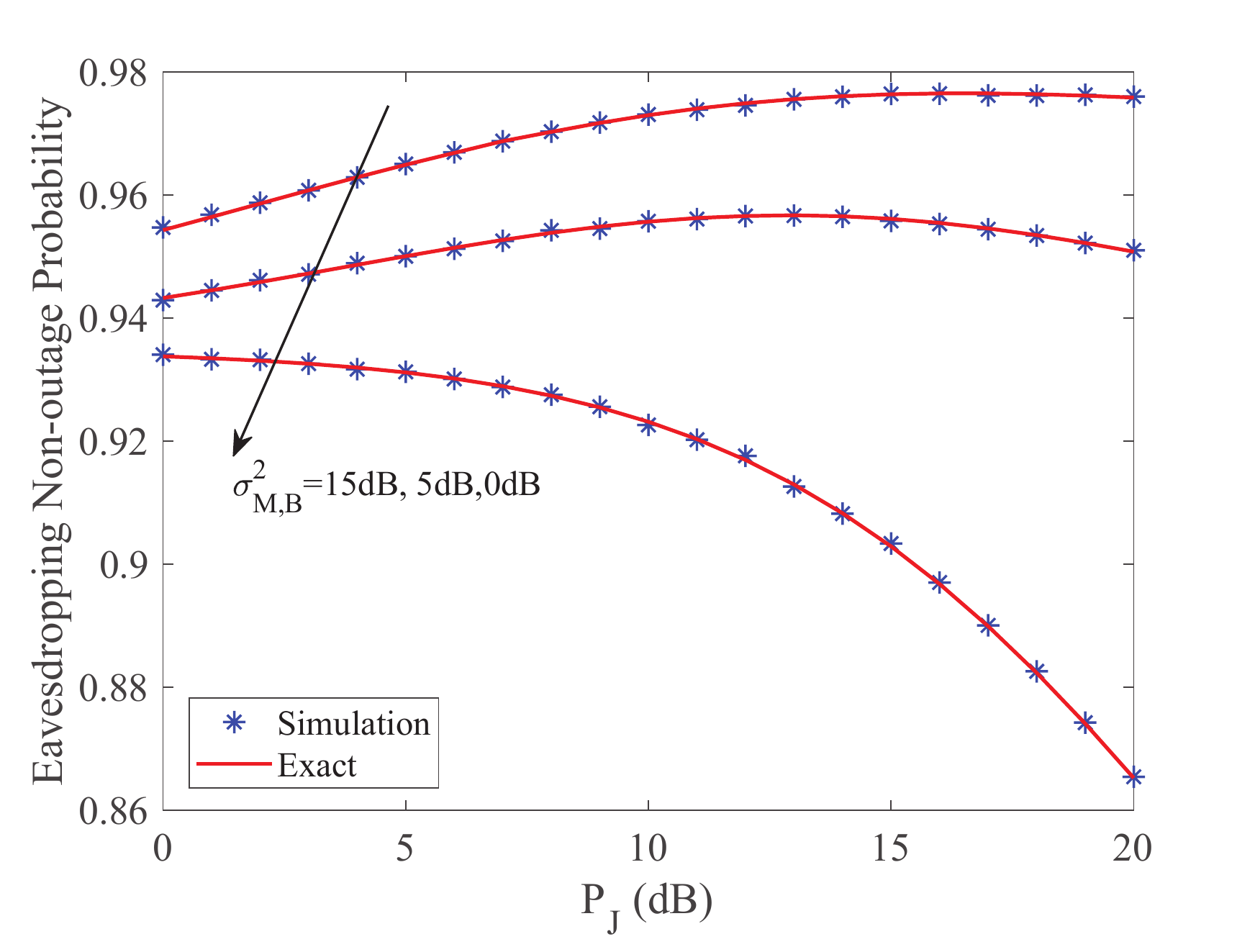}
\caption{Eavesdropping Non-outage Probability ${\bf{E}}\left[ \bf{X} \right] $ against the AN power $P_J$ without covert constraint.} \label{Fig2}
\end{figure}

 \begin{algorithm}
\caption{$E\left[ {\bf{X}} \right]$ maximization in different scenarios}
\begin{algorithmic}[1]
\STATE {\bf{Initialization:} $\delta$ and $P_J^*$}
\STATE {According to Lemma 1, the average minimum DEP can be obtained}.
\STATE  {if \eqref{problemb} is satisfied,
           \eqref{value_range} can be obtained.\\
           else $E\left[ {\bf{X}} \right]=0$, break;
        }
        \STATE  {Then, set $i=0$, $P_J=0$, $P_J^*=0$ and $j=0$.  \\
        \STATE {\bf{Repeat:}}\\
        a) $i=i+1$, $P_J=P_J+\Delta {P_J}$\\
        b) substitute $P_J$ into \eqref{ENP_exact}, we can obtain $E\left[ {\bf{X}} \right]$.\\
        c){if $E\left[ {\bf{X}} \right]\ge j$. Then
        $j$=$E\left[ {\bf{X}} \right]$ and
        $P_J^*=P_J$\\}
        \STATE {\bf{Until}} $P_J$ cannot satisfy \eqref{value_range}. The optimal AN power is ${P_{{J}}^*}$, and the maximum eavesdropping non-outage probability is $j$.
        }

\end{algorithmic}
\end{algorithm}
\section{SPECIAL CASES}
The covert surveillance model proposed in Section III can be useful for the general channel uncertainty cases, where we assume $0 < {\rho _{m,b}} < 1$ and $0 < \rho_{a,b}< 1$. In the following, we discuss the covert surveillance performance under several special cases.

\subsection{Special Case 1: $\rho_{a,b}=0$, $0 \le {\rho _{m,b}} \le 1$}
For this special case,  Bob has perfect knowledge of the instantaneous $h_{A,B}$, and the average power received at Bob is
 \begin{align}\label{average_power_sp1}\
{T_B} = \left\{ {\begin{array}{*{20}{c}}
{{P_a}{{\left| {{{h}_{A,B}}} \right|}^2} + \sigma _b^2}&{{H_0}}\\
{{P_a}{{\left| {{{h}_{A,B}}} \right|}^2} +  X_2 + {P_J}{{\left| {{{\widetilde h}_{M,B}}} \right|}^2} + \sigma _b^2}&{{H_1},}
\end{array}} \right.
\end{align}
where the average received power of $H_0$ is perfectly known by Bob. Since all channels are subject to quasi-static channel fading, the average power received at Bob of $H_1$ is higher than that of $H_0$ when Monitor injects the AN.
 If Bob already knows the average received power of $H_0$, he can determine the detection threshold as ${\Gamma^*  = {P_a}{{\left| {{h_{A,B}}} \right|}^2} + \sigma _b^2}$. Once Monitor injects the AN, it will be detected by Bob. Hence, it is better for Monitor to remain silent and $P_J^*=0$. Following that, \eqref{ENP} can be rewritten as
\begin{align}\label{ENP_sc1}\
{\bf{E}}\left[ \bf{X} \right] &= Prob\left( {\frac{{{{\left| {{h_{A,M}}} \right|}^2}}}{{\sigma _m^2}} \ge \frac{{{{\left| {{h_{A,B}}} \right|}^2}}}{{\sigma _b^2}}} \right)\nonumber\\
 &= \frac{{{\sigma ^2}_{A,M}{\sigma ^2}_b}}{{{\sigma ^2}_{A,M}{\sigma ^2}_b + {\sigma ^2}_{A,B}{\sigma ^2}_m}}.
\end{align}

\subsection{Special Case 2:  ${\rho _{a,b}} \ne 0$, $\rho_{m,b}=0$}
For this case, Bob has perfect knowledge of the Jamming Link and imperfect knowledge of the Suspicious Link. Bob's optimal detection threshold ${\Gamma ^ * }$ and minimum DEP ${{\xi ^*}} $ are, respectively, given by the following lemma.

{\emph{\textbf{Lemma 4:}}} For special case 2, ${\Gamma ^*}$ and ${\xi ^*}$ at Bob are given by
\begin{align}\label{optimal_menxian_sc3}\
{\Gamma ^*} = {X_1} + {{P_J}{{\left| {{h_{M,B}}} \right|}^2}} + \sigma _b^2
\end{align}
and
\begin{align}\label{optimal_DEP_sc3}\
{\xi ^*} =\exp \left( { - \frac{{{P_J}{{\left| {{h_{M,B}}} \right|}^2}}}{{{\rho _{a,b}}{P_a}{\sigma ^2}_{A,B}}}} \right).
\end{align}

{\emph{Proof:}}
According to \eqref{tnh0h2}, the average power received at Bob is
 \begin{align}\label{average_power_sc3}\
{T_B} = \left\{ {\begin{array}{*{20}{c}}
{X_1 + {P_a}{{\left| {{{\widetilde h}_{A,{B}}}} \right|}^2} + \sigma _b^2}&{{H_0}}\\
{X_1 + {P_a}{{\left| {{{\widetilde h}_{A,{B}}}} \right|}^2} + {{P_J}{{\left| {{h_{M,B}}} \right|}^2}} + \sigma _b^2}&{{H_1},}
\end{array}} \right.
\end{align}
where $X_1$ and ${{P_J}{{\left| {{h_{M,B}}} \right|}^2}}$ are perfectly known by Bob. Thus, the probabilities of false alarm and missed detection are, respectively, given by

\begin{align}\label{PFA_sc3}\
{P_{FA}} &= P\left( {{P_a}{{\left| {{{\widetilde h}_{A,B}}} \right|}^2} > \Gamma  - \sigma _b^2 - {X_1}} \right)\nonumber\\
 &= \left\{ {\begin{array}{*{20}{c}}
1&{\Gamma  \le {X_1} + \sigma _b^2}\\
{\exp \left( { - \frac{{\Gamma  - \sigma _b^2 - {X_1}}}{{{\rho _{a,b}}{P_a}\sigma _{A,B}^2}}} \right)}&{\Gamma  > {X_1} + \sigma _b^2}
\end{array}} \right.
\end{align}
and
\begin{align}\label{PMD_sc3}\
{P_{MD}} &= P\left( {{P_a}{{\left| {{{\widetilde h}_{A,B}}} \right|}^2} < \Gamma  - \sigma _b^2 - {X_1} - {P_J}{{\left| {{h_{M,B}}} \right|}^2}} \right)\nonumber\\
 &= \left\{ {\begin{array}{*{20}{c}}
0&{\Gamma  \le {X_1} + {P_J}{{\left| {{h_{M,B}}} \right|}^2} + \sigma _b^2}\\
{1 - \exp \left( { - \frac{{\Gamma  - \sigma _b^2 - {X_1} - {P_J}{{\left| {{h_{M,B}}} \right|}^2}}}{{{\rho _{a,b}}{P_a}\sigma _{A,B}^2}}} \right)}&{\Gamma  > {X_1} + {P_J}{{\left| {{h_{M,B}}} \right|}^2} + \sigma _b^2.}
\end{array}} \right.
\end{align}

Since $\xi  = {P_{FA}} + {P_{MD}}$, we can get the corresponding DEP as follows

\begin{align}\label{DEP_sc3}\
\begin{array}{l}
\xi  = \left\{ {\begin{array}{*{20}{c}}
1&{\Gamma  < \sigma _b^2 + {X_1}}\\
{{P_{FA}}}&{\sigma _b^2 + {X_1} \le \Gamma  < \sigma _b^2 + {X_1} + {{P_J}{{\left| {{h_{M,B}}} \right|}^2}}}\\
{{P_{FA}} + {P_{MD}}}&{\sigma _b^2 + {X_1} + {{P_J}{{\left| {{h_{M,B}}} \right|}^2}} \le \Gamma }.
\end{array}} \right.
\end{array}
\end{align}

Clearly, when ${X_1} + \sigma _b^2 \le \Gamma  < {X_1} + {{P_J}{{\left| {{h_{M,B}}} \right|}^2}}+ \sigma _b^2$, $\xi$ decreases with $\Gamma$. When ${X_1} + {{P_J}{{\left| {{h_{M,B}}} \right|}^2}} + \sigma _b^2 \le \Gamma $, $\xi $ increases with $\Gamma$. Hence, ${\Gamma ^*} = {X_1} +{{P_J}{{\left| {{h_{M,B}}} \right|}^2}}+ \sigma _b^2$, and ${\xi ^*} =\exp \left( { - \frac{{{P_J}{{\left| {{h_{M,B}}} \right|}^2}}}{{{\rho _{a,b}}{P_a}{\sigma ^2}_{A,B}}}} \right)$. The proof of this lemma is completed.

Then, the average minimum DEP can be calculated as
\begin{align}\label{Special_F}\
\overline {{\xi ^*}}  &= \int_{\rm{0}}^\infty  {\exp{\left(-\frac {{{P_J}{{\left| {{h_{M,B}}} \right|}^2}} } {{\rho _{a,b}}{P_a}{\sigma ^2}_{A,B}} \right)} \times {f_{{{P_J}{{\left| {{h_{M,B}}} \right|}^2}} }}\left( x \right)}dx \nonumber\\
 &= \frac{{{\rho _{a,b}}{P_a}{\sigma ^2}_{A,B}}}{{{\rho _{a,b}}{P_a}{\sigma ^2}_{A,B} + {P_J}{\sigma ^2}_{M,B}}},
\end{align}
where ${f_{{{{P_J}{{\left| {{h_{M,B}}} \right|}^2}}}}}\left( x \right) = \frac{1}{{{P_J}\sigma _{M,B}^2}}\exp \left( { - \frac{x}{{{P_J}\sigma _{M,B}^2}}} \right)$ is the PDF of ${{P_J}{{\left| {{h_{M,B}}} \right|}^2}}$.

{\emph{Remark 2:}}
When compared to special case $1$,  Bob's imperfect knowledge of $h_{A,B}$ is more crucial to covert surveillance. Once Bob has perfect knowledge of $h_{A,B}$, Monitor cannot inject AN power at all. However, when Bob has perfect knowledge of $h_{M,B}$,
Monitor still can inject AN to assist eavesdropping under the covert constraint.
\subsection{Special Case 3: $\rho_{a,b}=1$, $0 \le {\rho _{m,b}} \le 1$}
According to $Remark$ $1$, Bob's AN detection ability degrades with $\rho_{a,b}$, and has non-monotonic relationship with $\rho_{m,b}$. Hence, the most desirable case for Monitor (i.e, $\overline {{\xi ^*}}$  is maximized.) is $\rho_{a,b}=1$. In this case, Bob's optimal detection threshold ${\Gamma ^ * }$ and minimum DEP $\overline {{\xi ^*}} $ under this special case are, respectively, given by the following lemma.

{\emph{\textbf{Lemma 3:}}} For special case 3, ${\Gamma ^ * }$ and $\overline {{\xi ^*}}$ at Bob are, respectively, given by
\begin{align}\label{optimal_menxian_sc2}\
{\Gamma ^*} = {X_2} + \sigma _b^2 + {k_2}
\end{align}
and
\begin{align}\label{optimal_DEP_sc2}\
{\xi ^*} &= 1 + \exp \left( { - \frac{{{X_2} + {k_2}}}{{{P_a}\sigma _{A,B}^2}}} \right)\nonumber\\
 &- \frac{{{P_a}\sigma _{A,B}^2\exp \left( { - \frac{{{k_2}}}{{{P_a}\sigma _{A,B}^2}}} \right) - {\rho _{m,b}}{P_J}\sigma _{M,B}^2\exp \left( { - \frac{{{k_2}}}{{{\rho _{m,b}}{P_J}\sigma _{M,B}^2}}} \right)}}{{{P_a}\sigma _{A,B}^2 - {\rho _{m,b}}{P_J}\sigma _{M,B}^2}},
\end{align}
where ${k_2} = \frac{{{\rho _{m,b}}{P_J}\sigma _{M,B}^2{P_a}\sigma _{A,B}^2}}{{{\rho _{m,b}}{P_J}\sigma _{M,B}^2 - {P_a}\sigma _{A,B}^2}}\ln \left( {1 - \left( {1 - \frac{{{\rho _{m,b}}{P_J}\sigma _{M,B}^2}}{{{P_a}\sigma _{A,B}^2}}} \right)\exp \left( { - \frac{{{X_2}}}{{{P_a}\sigma _{A,B}^2}}} \right)} \right)$.

{\emph{Proof:}}
According to \eqref{tnh0h2}, the average power received at Bob is
 \begin{align}\label{average_power_sc2}\
{T_B} = \left\{ {\begin{array}{*{20}{c}}
{{P_a}{{\left| {{h_{A,B}}} \right|}^2} + \sigma _b^2}&{{H_0}}\\
{{P_a}{{\left| {{h_{A,B}}} \right|}^2} + X_2 + {P_J}{{\left| {{{\widetilde h}_{M,B}}} \right|}^2} + \sigma _b^2}&{{H_1},}
\end{array}} \right.
\end{align}
where $X_2$ is perfectly known by Bob. Thus, the probabilities of false alarm and missed detection are, respectively, given by

\begin{align}\label{PFA_sc2}\
{P_{FA}} &= P\left( {{P_a}{{\left| {{h_{A,B}}} \right|}^2} > \Gamma  - \sigma _b^2} \right)\nonumber\\
& = \left\{ {\begin{array}{*{20}{c}}
1&{\Gamma  \le \sigma _b^2}\\
{\exp \left( { - \frac{{\Gamma  - \sigma _b^2}}{{{P_a}\sigma _{A,B}^2}}} \right)}&{\Gamma  > \sigma _b^2}
\end{array}} \right.
\end{align}
and
\begin{align}\label{PMD_sc2}\
{P_{MD}} &= P\left( {{P_a}{{\left| {{h_{A,B}}} \right|}^2} + {P_J}{{\left| {{{\widetilde h}_{M,B}}} \right|}^2} < \Gamma  - \sigma _b^2 - {X_2}} \right)\nonumber\\
 &= \left\{ {\begin{array}{*{20}{c}}
0&{\Gamma  \le {X_2} + \sigma _b^2}\\
{1 - \frac{{{P_a}\sigma _{A,B}^2\exp \left( { - \frac{{\Gamma  - \sigma _b^2 - {X_2}}}{{{P_a}\sigma _{A,B}^2}}} \right) - {\rho _{m,b}}{P_J}\sigma _{M,B}^2\exp \left( { - \frac{{\Gamma  - \sigma _b^2 - {X_2}}}{{{\rho _{m,b}}{P_J}\sigma _{M,B}^2}}} \right)}}{{{P_a}\sigma _{A,B}^2 - {\rho _{m,b}}{P_J}\sigma _{M,B}^2}}}&{\Gamma  > {X_2} + \sigma _b^2.}
\end{array}} \right.
\end{align}
Since $\xi  = {P_{FA}} + {P_{MD}}$, we can get the corresponding DEP as
 \begin{align}\label{DEP_sc2}\
\xi  = \left\{ {\begin{array}{*{20}{c}}
1&{\Gamma  <\sigma _b^2}\\
{{P_{FA}}}&{ \sigma _b^2 \le \Gamma  < {X_2} + \sigma _b^2}\\
{{P_{FA}} + {P_{MD}}}&{{X_2} + \sigma _b^2 \le \Gamma }.
\end{array}} \right.
\end{align}
It is obvious that $\xi$ decreases with $\Gamma$ when ${ \sigma _b^2 \le \Gamma  < {X_2} + \sigma _b^2}$. Similar to Lemma 1, we can prove that $\xi$ decreases with $\Gamma$ when ${X_2}+\sigma _b^2\le\Gamma<{X_2}+k_2 + \sigma _b^2$, and $\xi$ increases with $\Gamma$ when $\Gamma>{X_2}+k_2 + \sigma _b^2$.
Hence, when ${\Gamma ^*} = {X_2}+k_2 + \sigma _b^2 $, $\xi$ can achieve the minimum value, respectively. Thus, the average minimum DEP $\overline {{\xi ^*}} $ can be calculated as
\begin{align}\label{optimal_DEP_sc2}\
\overline {{\xi ^*}}  &= \int_0^\infty  {{\xi ^*}{f_{{X_2}}}\left( x \right)} dx\nonumber\\
 &= 1 - {\left( {1 - \frac{{{\rho _{m,b}}{P_J}\sigma _{M,B}^2}}{{{P_a}\sigma _{A,B}^2}}} \right)^{ - \frac{{{P_a}\sigma _{A,B}^2}}{{\left( {1 - {\rho _{m,b}}} \right){P_J}\sigma _{M,B}^2}}}}\frac{{{P_a}\sigma _{A,B}^2}}{{\left( {1 - {\rho _{m,b}}} \right){P_J}\sigma _{M,B}^2}}\nonumber\\
 &\times \int_{\frac{{{\rho _{m,b}}{P_J}\sigma _{M,B}^2}}{{{P_a}\sigma _{A,B}^2}}}^1 {{x^{{{\left( {1 - \frac{{{\rho _{m,b}}{P_J}\sigma _{M,B}^2}}{{{P_a}\sigma _{A,B}^2}}} \right)}^{ - 1}}}}{{\left( {1 - x} \right)}^{\frac{{{P_a}\sigma _{A,B}^2}}{{\left( {1 - {\rho _{m,b}}} \right){P_J}\sigma _{M,B}^2}} - 1}}} dx,
\end{align}
the exactness of \eqref{optimal_DEP_sc2} is verified in Fig. 4. Similar to Appendix B, we can verify that $\rho_{m,b}$ has a non-monotonic effect on $\overline {{\xi ^*}} $ in this case. Hence, the most desirable case is $\rho_{a,b}=1$ and $\rho_{m,b}=1$ or $\rho_{a,b}=1$ and $\rho_{m,b}=0$.


In particular, when $\rho_{m,b}=1$, ${\Gamma ^ * }$ and $ \overline{{\xi ^*}}$ at Bob are, respectively, given by
\begin{align}\label{optimal_menxian_sc2_1}\
\Gamma^*  = {k_3} + \sigma _b^2
\end{align}
and
\begin{align}\label{optimal_DEP_sc2_1}\
\overline {{\xi ^*}} & = \exp \left( { - \frac{{{k_3}}}{{{P_a}\sigma _{A,B}^2}}} \right)\nonumber\\
 &+ 1 - \frac{1}{{{P_a}\sigma _{A,B}^2 - {P_J}\sigma _{M,B}^2}}\left( {{P_a}\sigma _{A,B}^2\exp \left( { - \frac{{{k_3}}}{{{P_a}\sigma _{A,B}^2}}} \right) - {P_J}\sigma _{M,B}^2\exp \left( { - \frac{{{k_3}}}{{{P_J}\sigma _{M,B}^2}}} \right)} \right),
\end{align}
where ${k_3} = \frac{{{P_J}\sigma _{M,B}^2{P_a}\sigma _{A,B}^2}}{{{P_J}\sigma _{M,B}^2 - {P_a}\sigma _{A,B}^2}}\ln \frac{{{P_J}\sigma _{M,B}^2}}{{{P_a}\sigma _{A,B}^2}}$. Also, when $\rho_{m,b}=0$, ${\Gamma ^ * }$ and $ {{\xi ^*}} $ at Bob are, respectively, given by
\begin{align}\label{optimal_menxian_sc2_2}\
{\Gamma ^*} = {P_J}{\left| {{h_{M,B}}} \right|^2} + \sigma _b^2
\end{align}
and
\begin{align}\label{optimal_DEP_sc2_2}\
\overline{{\xi ^*}} = \frac{1}{{1 + \frac{{{P_J}\sigma _{M,B}^2}}{{{P_a}\sigma _{A,B}^2}}}}.
\end{align}

Note that we investigate special case 3 depending on $\rho_{a,b}=1$ and the different values of $\rho_{m,b}$. To provide more insight, Fig. 4 shows $\overline {{\xi ^*}}$ at a  given $\rho_{m,b}$ versus the variable $\rho_{a,b}$ (in the range between $0$ and $1$). It is observed that $\overline {{\xi ^*}}$ increases as a function of $\rho_{a,b}$. Note that $\overline {{\xi ^*}}=0$ when $\rho_{a,b}=0$, which is verified in special case 1.

\begin{figure}[ht]
\centering
\includegraphics[width=0.5\textwidth]{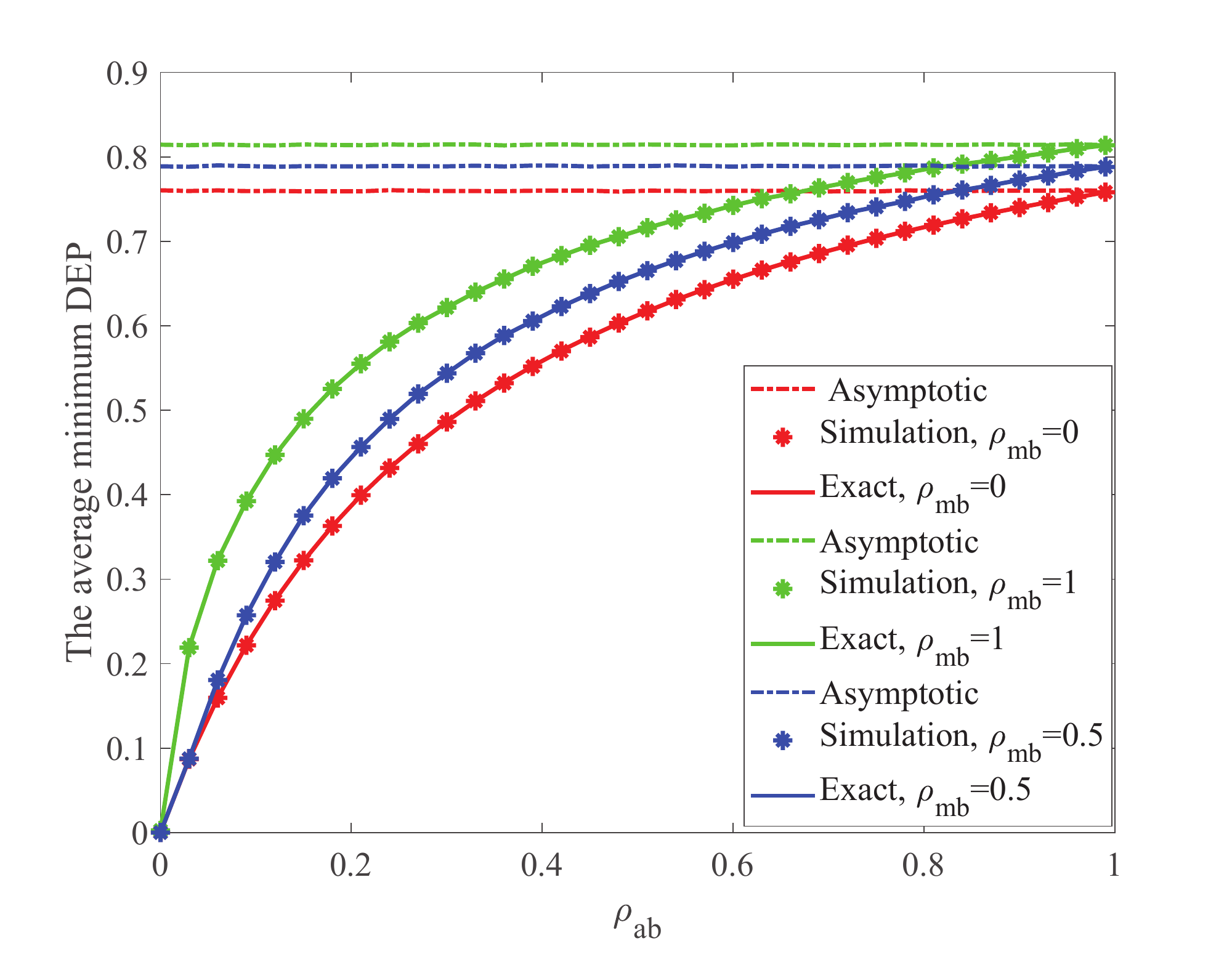}
\caption{The average minimum DEP $\overline {{\xi ^*}} $ against $\rho_{a,b}$ with different values of $\rho_{m,b}$.} \label{Fig3}
\end{figure}

\section{Numerical Results}
In this section, numerical results for covert surveillance performance are presented under channel uncertainties for both the Suspicious link and the Jamming link. Assume the noise variance at
Bob $\sigma^2_{b}=1$ and the noise variance at Monitor $\sigma^2_{m}=1$. The self-interference coefficient is $\eta=0.5$. The transmission power $P_{a}$, $P_{J}$  and $P_J^{max}$ are, respectively, normalized by the noise variance, denoted as $\bar{P}_{a}$, ${\bar{P}}_J$  and ${\bar{P}}_J^{\max}$ in the following simulation,  i.e., ${\bar{P}}_{a}=P_{a}/{\sigma^2_{b}}$, ${\bar{P}}_J={{P}}_J/{\sigma^2_{b}}$, and ${\bar{P}}_J^{\max}={{P}}_J^{\max}/{\sigma^2_{b}}$. In addition, the channel qualities of Alice $\rightarrow$ Bob,  Alice $\rightarrow$ Monitor,  Monitor $\rightarrow$ Bob, and Monitor $\rightarrow$ Monitor are normalized as ${\bar\gamma_{A,B}}={\sigma^2_{A,B}}/{\sigma^2_{b}}$, ${\bar\gamma_{A,M}}={\sigma^2_{A,M}}/{\sigma^2_{m}}$, ${\bar\gamma_{M,B}}={\sigma^2_{M,B}}/{\sigma^2_{b}}$, and ${\bar\gamma_{M,M}}={\sigma^2_{M,M}}/{\sigma^2_{m}}$, respectively. Unless otherwise specified, ${\bar{P}}_{a}=25dB$, and ${\bar{P}}_J^{\max}=25dB$.

For comparison, the performance of the two benchmark schemes proposed in \cite{7321779}  are also investigated as follows: 1) Passive eavesdropping, i.e., $P_J^*=0$. 2) Proactive eavesdropping with constant AN power, i.e., $P_J^*=P_J^{\max }$. However, due to the covert constraint, Monitor can not inject AN with full power $P_J^{\max }$. Therefore, we improve the second scheme into a proactive eavesdropping with constant covert AN power scheme i.e., $P_J^*=\min \left( {P_J^{\max },P_J^{covert}} \right)$.

Figure 5 presents the average minimum DEP $\overline {{\xi ^*}} $ versus $\rho_{a,b}$ and $\rho_{m,b}$ for different $P_J$. It can be seen from Fig. 5(a) and Fig. 5(b) that increasing $\rho_{a,b}$ leads to the degradation of  Bob's detection performance. In contrast, $\rho_{m,b}$ causes non-monotonic influence on $\overline {{\xi ^*}} $ just as shown in Fig. 5(b), which is consistent with $Remark$ $1
 $. Specifically, when $\rho_{a,b}=0$ and $\rho_{m,b}\ne0$, we obtain $\overline {{\xi ^*}}=0$. However, when $\rho_{m,b}=0$ and $\rho_{a,b}\ne0$, we obtain $\overline {{\xi ^*}}\ne0$. This means that Bob's imperfect knowledge of $h_{A,B}$ is crucial to covert surveillance. If Bob has perfect knowledge of $h_{A,B}$, Monitor cannot inject AN power at all. Nevertheless, when Bob has perfect knowledge of $h_{M,B}$, Monitor still can inject AN power with a non-unity probability of being detected by Bob. Note that the value of $\overline {{\xi ^*}} $ is maximized when $\rho_{a,b}=1$, which can be regarded as an upper-bound for Monitor's covert AN transmission, as given in special case 3.

\begin{figure}
\centering
\subfigure[${\bar{P}}_J=10dB$]{\includegraphics[width=0.48\textwidth]{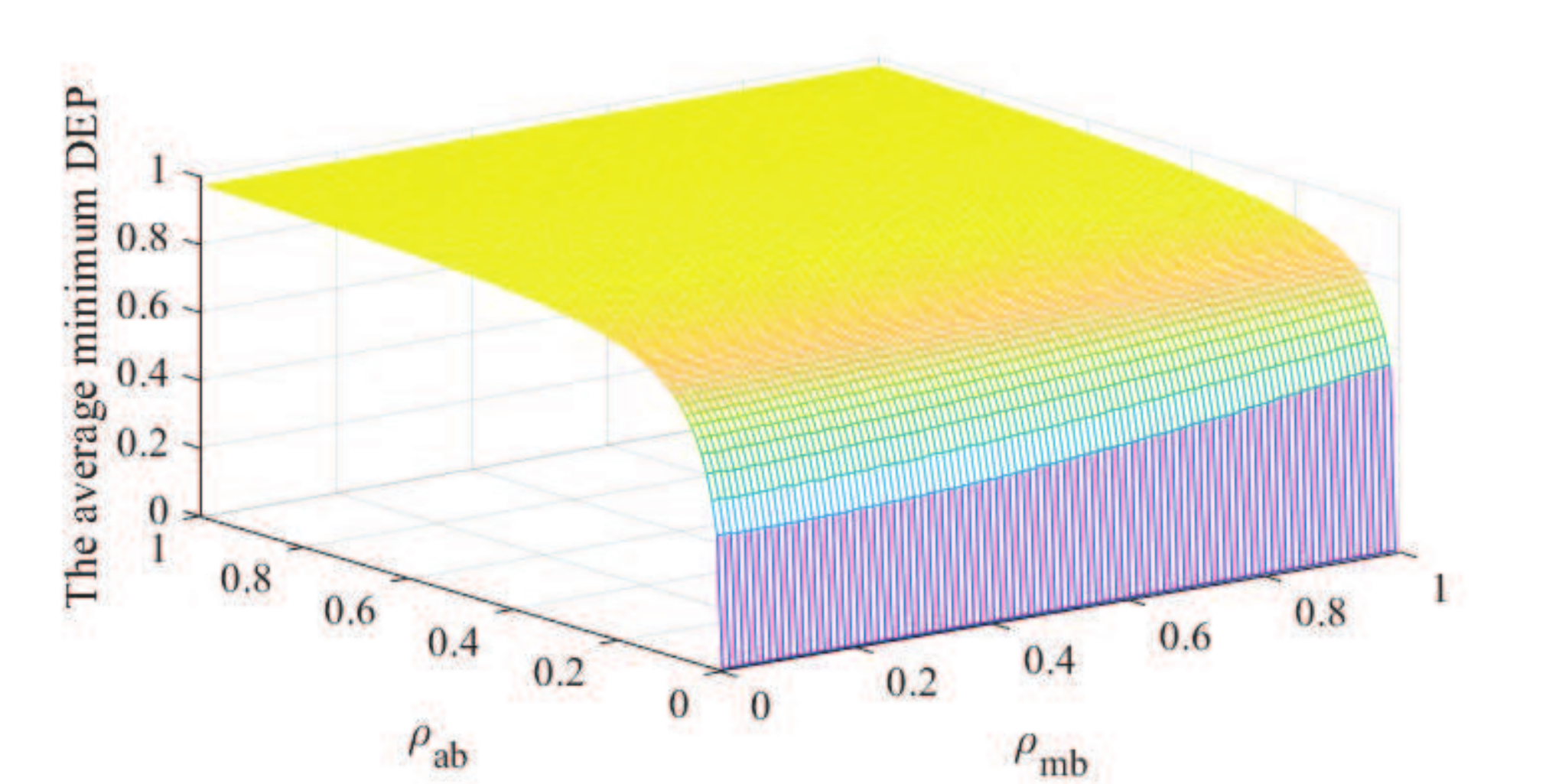}}
\subfigure[${\bar{P}}_J=25dB$]{\includegraphics[width=0.48\textwidth]{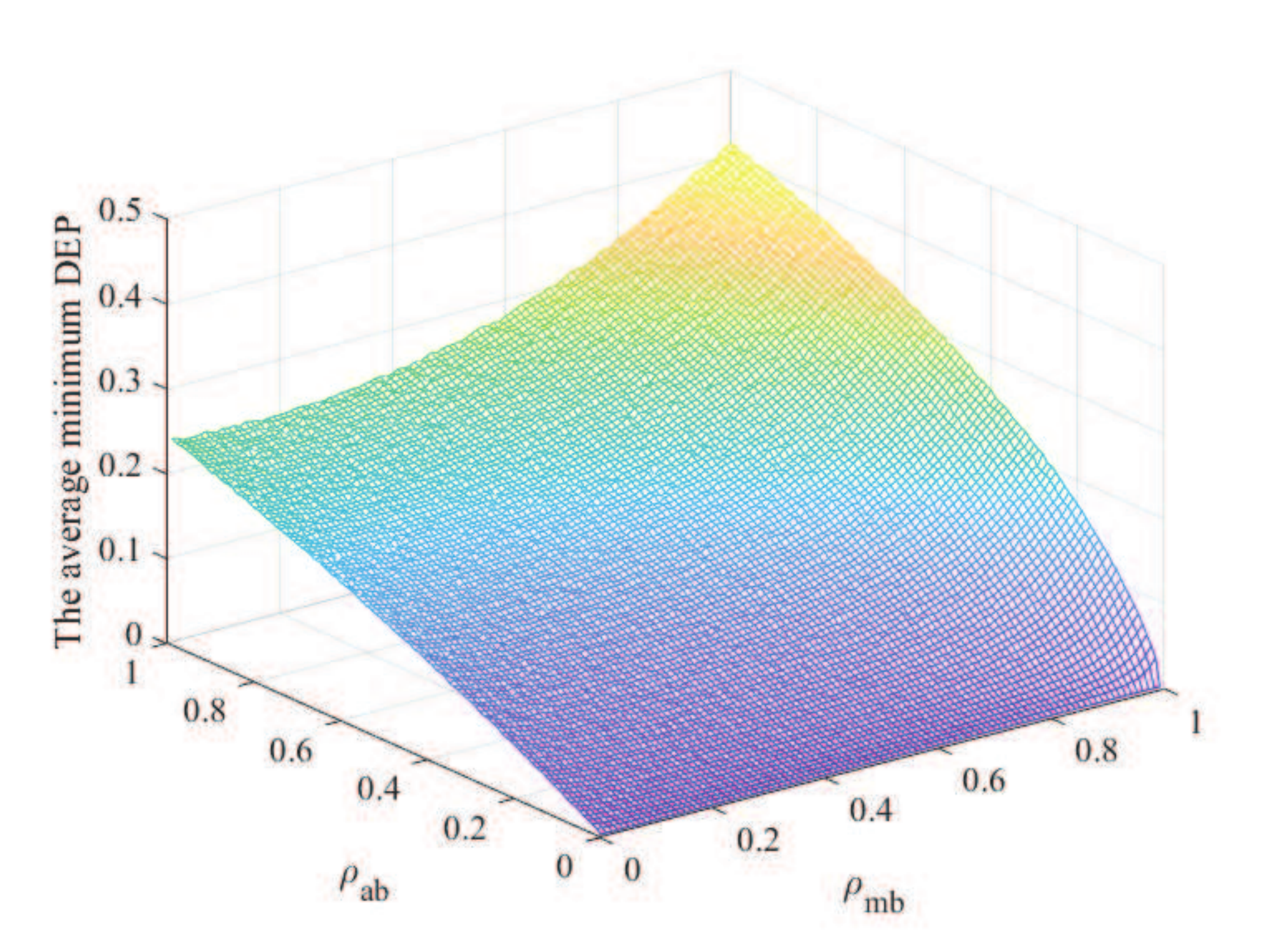}}
\caption{The average minimum DEP $\overline {{\xi ^*}} $ against $\rho_{m,b}$. $\bar\gamma_{A,B}=1$ and $\bar\gamma_{A,M}=1$.} \label{Fig5}
\label{fig6}
\end{figure}


\begin{figure}[ht]
\centering
\includegraphics[width=0.8\textwidth]{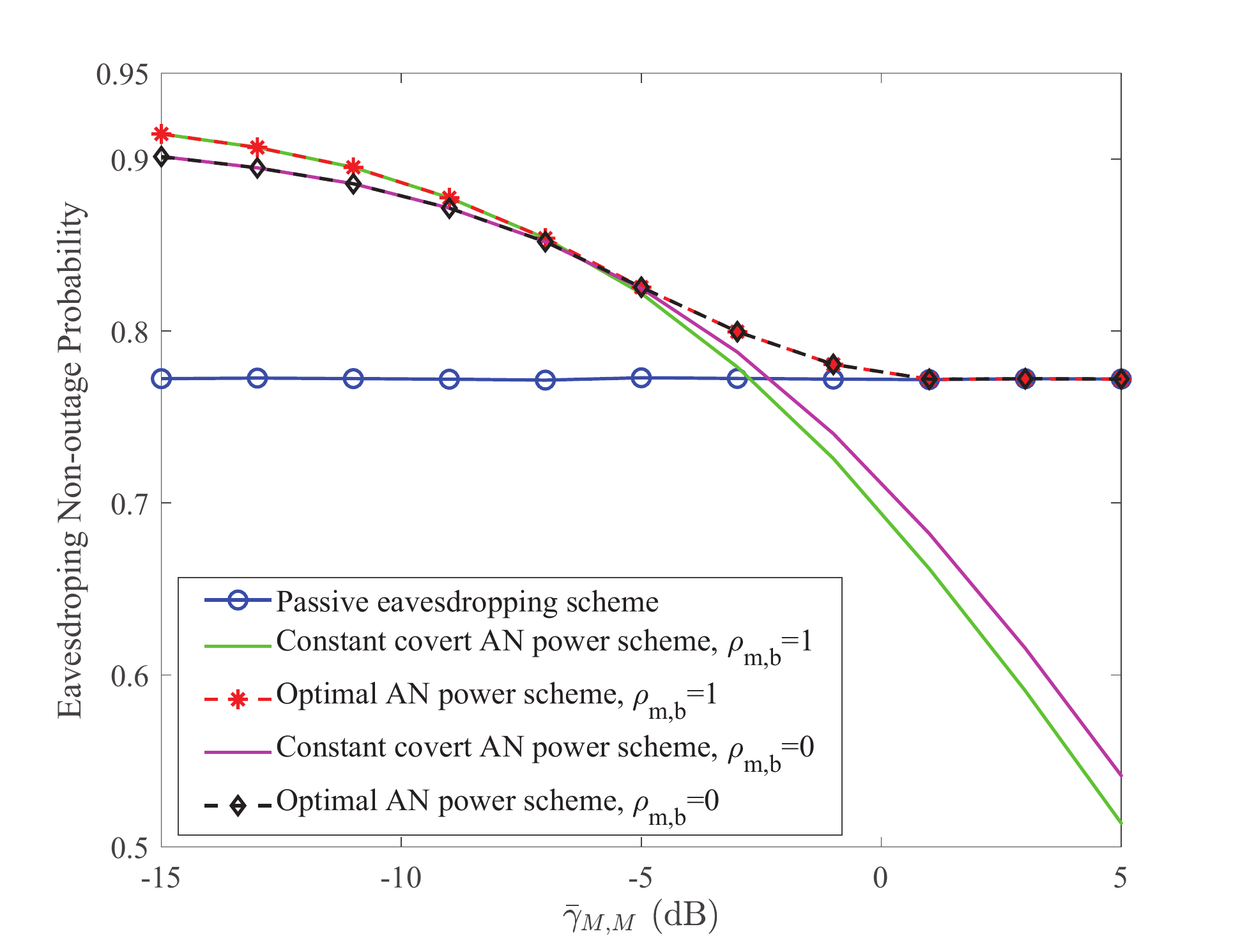}
\caption{Eavesdropping Non-outage Probability ${\bf{E}}\left[ \bf{X} \right] $ against $\bar\gamma_{M,M}$. $\rho_{a,b}=0.5$, $\rho_{m,b}=0.5$, $\eta=0.1$, $\delta=0.5$, $\bar\gamma_{A,B}=5dB$, $\bar\gamma_{M,B}=10dB$, $\bar\gamma_{A,M}=-10dB$, $\sigma^2_{b}=1$, and $\sigma^2_{m}=1$.} \label{Fig7}
\end{figure}

Figure 6 shows eavesdropping non-outage probability ${\bf{E}}\left[ \bf{X} \right] $ versus $\bar\gamma_{M,M}$. It is clear that $\bar\gamma_{M,M}$ has no impact on ${\bf{E}}\left[ \bf{X} \right] $ of the passive eavesdropping scheme where $P_J=0$. In contrast, for the constant covert AN power scheme under the covert constraint, ${\bf{E}}\left[ \bf{X}\right] $ decreases with $\bar\gamma_{M,M}$ due to the severe self-interference at Monitor.
Specifically, when $\bar\gamma_{M,M} < -7dB$, the AN power introduces a higher level of interference power at Bob than the self-interference power at Monitor.
Hence, it is better to inject AN towards Bob with $P_J^*=\min \left( {P_J^{\max },P_J^{covert}} \right)$. Thus, the curves of optimal AN power scheme coincide with the curves of constant covert AN power scheme. When $-7dB<\bar\gamma_{M,M} < 2dB$, the optimal AN power $P_J^*$ lies in the range $\left( {0,\min \left( {P_J^{\max },P_J^{covert}} \right)} \right)$. It is observed that the optimal AN power scheme achieves higher eavesdropping performance, compared to the two benchmark schemes.
When $\bar\gamma_{M,M}>2dB$, due to the severe self-interference, Monitor keeps silent and $P_J^*=0$. Then, the curve of optimal AN power scheme coincides with the curve of passive eavesdropping scheme. As expected, the proposed scheme substantially outperforms the two benchmark schemes regardless of $\bar\gamma_{M,M}$. Moreover, it is better for Monitor to remain silent when self-interference becomes severe. Hence, we find that $\rho_{m,b}$ has no effect on ${\bf{E}}\left[ \bf{X} \right] $ when $\bar\gamma_{M,M}>2dB$. We conclude that increasing the channel uncertainty of the Jamming Link is an effective means to enhance covert surveillance
performance only when the self-interference at Monitor is small.

\begin{figure}[ht]
\centering
\includegraphics[width=5in]{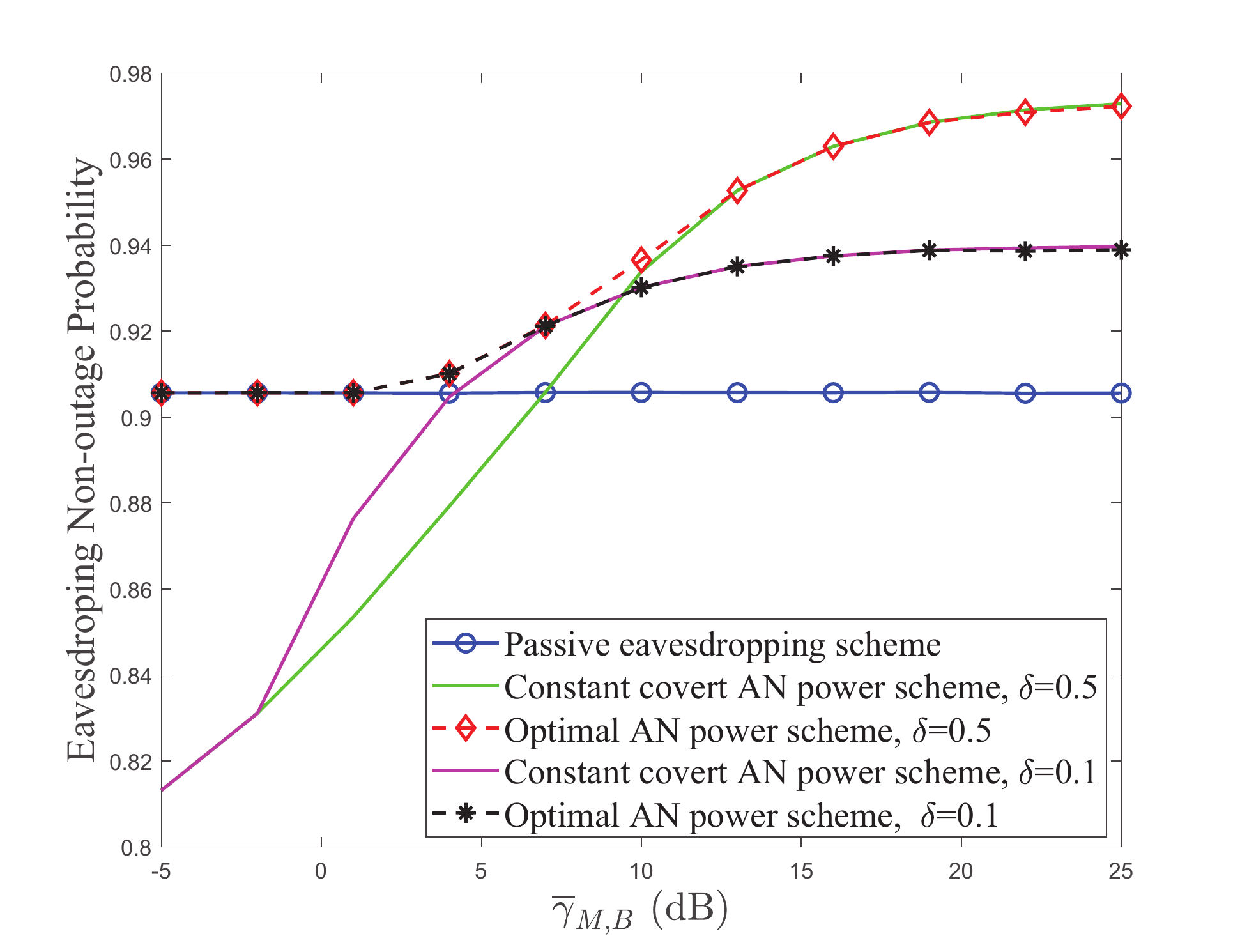}
\caption{Eavesdropping Non-outage Probability ${\bf{E}}\left[ \bf{X} \right] $ against $\bar\gamma_{M,B}$. $\rho_{a,b}=0.5$, $\rho_{m,b}=0.5$, $\eta=0.1$, $\bar\gamma_{A,B}=10dB$, $\bar\gamma_{M,M}=-10dB$, $\bar\gamma_{A,M}=-10dB$, $\sigma^2_{b}=1$, and $\sigma^2_{m}=1$.} \label{Fig8}
\end{figure}

${\bf{E}}\left[ \bf{X} \right] $ for different $\bar\gamma_{M,B}$ is illustrated in Fig. 7. Different from $\bar\gamma_{M,M}$ in Fig. 6, $\bar\gamma_{M,B}$ not only directly affects ${\bf{E}}\left[ \bf{X} \right] $ according to \eqref{ENP}, but also indirectly affects ${\bf{E}}\left[ \bf{X} \right] $ according to the covert constraint \eqref{DEP_average_minmum}. When $\bar\gamma_{M,B}<-2dB$, the curves of constant covert AN power scheme converge together for different values of $\delta$ since Monitor can inject AN with maximum power when $\bar\gamma_{M,B}$ is small. However, due to the fact that AN power introduces higher self-interference at Monitor, Monitor should keep silent and $P_J^*=0$. In this case, the curves of optimal AN power scheme coincide with the curves of the passive eavesdropping scheme. When $\bar\gamma_{M,B}=-2dB$, the curves of constant AN power scheme begin to separate, since the covert constraint $\delta=0.1$ is stricter than $\delta=0.5$. Moreover, when $\bar\gamma_{M,B}<7dB$, ${\bf{E}}\left[ \bf{X}\right]$ of the optimal AN power scheme has same value for different $\delta$. However, as $\bar\gamma_{M,B}$ continues to increase, the value of ${\bf{E}}\left[ \bf{X} \right]$ with $\delta=0.5$ is larger than that with $\delta=0.1$. This observation demonstrates that a loose covert constraint can achieve higher eavesdropping performance only when the AN power at Bob is larger than the self-interference power at Monitor. In other words, once the self-interference becomes severe, Monitor should remain silent. Thus, the covert constraint has no effect on the surveillance performance.

\begin{figure}
\centering
\subfigure[$\bar\gamma_{M,M}=-5dB$]{\includegraphics[width=0.48\textwidth]{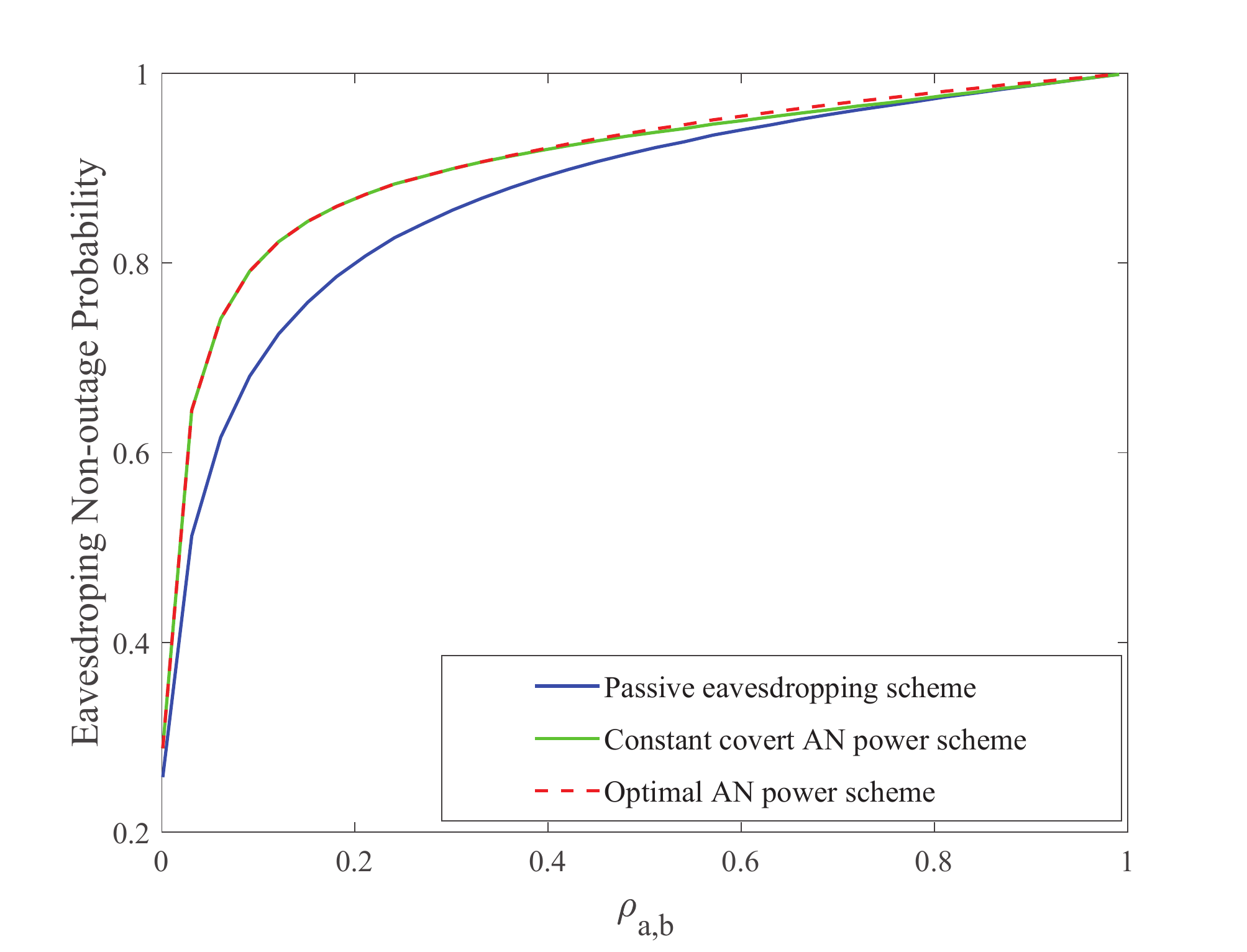}}
\subfigure[$\bar\gamma_{M,M}=-2dB$]{\includegraphics[width=0.48\textwidth]{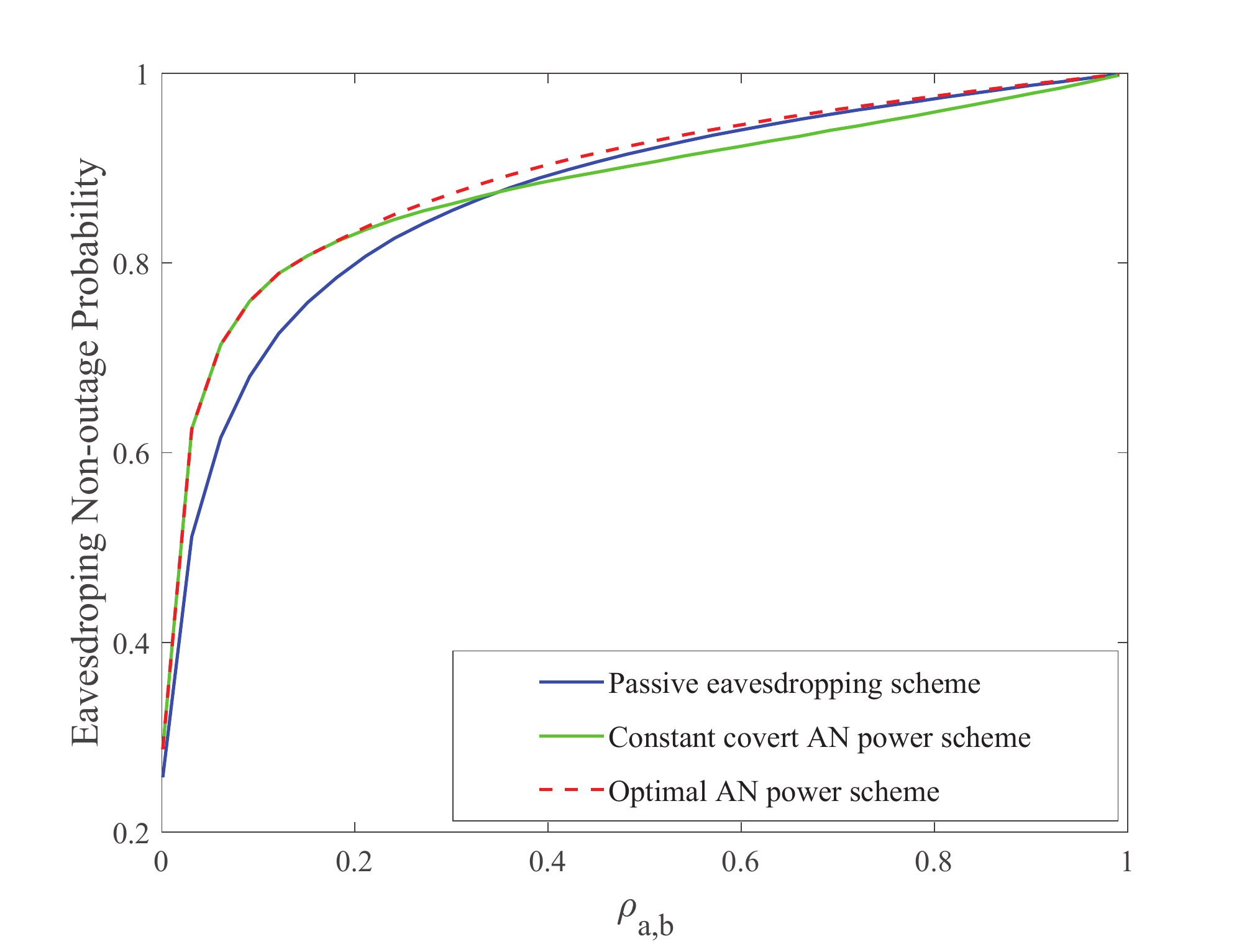}}
\caption{Eavesdropping Non-outage Probability ${\bf{E}}\left[ X \right] $ against $\rho_{a,b}$. $\rho_{a,b}=0.5$, $\rho_{m,b}=0.5$, $\eta=0.1$, $\delta=0.5$, $\bar\gamma_{A,B}=-5dB$, $\bar\gamma_{M,B}=1$, $\bar\gamma_{A,M}=-10dB$, $\sigma^2_{b}=1$, and $\sigma^2_{m}=1$.} \label{Fig9}
\end{figure}

Figure 8 investigates the eavesdropping non-outage probability ${\bf{E}}\left[ \bf{X} \right] $ versus $\rho_{a,b}$.
It is clear that the curve of optimal
AN power scheme coincides with the curve of constant covert AN power scheme when $\bar\gamma_{M,M}=-5dB$ in Fig. 8(a). This figure demonstrates that when the AN introduces a higher level of interference power at Bob than the self-interference power at Monitor,
it is better to use full AN power. However, when $\bar\gamma_{M,M}=-2dB$ in Fig. 8(b), the self-interference becomes severe. Therefore, it is better for Monitor to remain silent, and the curve of optimal AN power scheme coincides with the curve of passive eavesdropping scheme. In Fig. 8(b), we find that the curve of optimal AN power scheme firstly coincides with the curve of constant covert AN power scheme, and then coincides with the curve of passive eavesdropping scheme  as $\rho_{a,b}$ is increased, because $\rho_{a,b}$ affects the relationship between  ${\bf{E}}\left[ \bf{X} \right] $ and $P_J$  as shown in \eqref{P_J}. When the value of $\rho_{a,b}$ is small, ${\bf{E}}\left[ \bf{X} \right] $ increases with $P_J$. Hence, the constant covert AN power scheme can achieve better performance than the passive eavesdropping scheme. Nevertheless, when the value of $\rho_{a,b}$ is large, ${\bf{E}}\left[ \bf{X} \right] $ decreases with $P_J$ and the passive eavesdropping scheme outperforms the constant covert AN power scheme. In both Fig. 8(a) and Fig. 8(b), the proposed optimal AN power scheme achieves the best performance in terms of eavesdropping non-outage probability.  Moreover, we see that ${\bf{E}}\left[ \bf{X} \right]$ increases with $\rho_{a,b}$. It indicates that the channel uncertainty of the Suspicious link is an effective way to achieve a better eavesdropping performance. However, channel uncertainty of the Jamming link only has effect on eavesdropping performance when the self-interference at Monitor is small just as shown in Fig. 6.  Hence, we conclude that Suspicious Link uncertainty is crucial to improve the surveillance performance.


\section{Conclusion}
In this paper, we study the performance of wireless surveillance systems under a covert constraint. Considering
the channel uncertainties of both the Suspicious Link and Jamming Link, closed-form
expressions for the optimal AN detection threshold and the average minimum DEP are derived. Numerical results show that the channel uncertainty can effectively improve covert surveillance performance. Specifically, Suspicious Link channel uncertainty has greater influence than Jamming link channel uncertainty on surveillance performance. Once the suspicious user has perfect channel knowledge of the Suspicious Link, injecting AN by the legitimate Monitor does not assist the eavesdropping at all. However, even if the suspicious user has perfect knowledge of the Jamming Link, the legitimate Monitor can still inject AN under the covert constraint. In addition, Suspicious Link channel uncertainty can always affect covert surveillance performance. In contrast, Jamming Link channel uncertainty can affect the covert surveillance performance
only when the interference caused by AN at Bob is greater than the self-interference at the full-duplex Monitor.

We hope that this paper can provide a new paradigm for designing legitimate surveillance schemes when the suspicious users are intelligent and capable of detecting the AN from Monitor.

\begin{appendices}
\section{}

We can rewrite \eqref{DEP_average_minmum} as
\begin{align}\label{T_B1_T}\
\overline {{\xi ^*}}  = \int_{\rm{0}}^\infty  {\left( \begin{array}{l}
1 + \exp \left( { - \frac{{{P_J}\left( {1 - {\rho _{m,b}}} \right)x{\rm{ + }}{k_1}}}{{{\rho _{a,b}}{P_a}\sigma _{A,B}^2}}} \right)\\
 - \frac{{{\rho _{a,b}}{P_a}\sigma _{A,B}^2\exp \left( { - \frac{{{k_1}}}{{{\rho _{a,b}}{P_a}\sigma _{A,B}^2}}} \right) - {\rho _{m,b}}{P_J}\sigma _{M,B}^2\exp \left( { - \frac{{{k_1}}}{{{\rho _{m,b}}{P_J}\sigma _{M,B}^2}}} \right)}}{{{\rho _{a,b}}{P_a}\sigma _{A,B}^2 - {\rho _{m,b}}{P_J}\sigma _{M,B}^2}}
\end{array} \right) \times \varphi \left( x \right)} dx,
\end{align}
where $\varphi \left( x \right) = \frac{1}{{\sigma _{M,B}^2}}\exp \left( { - \frac{1}{{\sigma _{M,B}^2}}x} \right)$.

Then, since $P_J$ has no effect on $\varphi\left( x \right)$, the derivative $\frac{{d\overline {{\xi ^*}} }}{{d{P_J}}}$ can be written as
\begin{align}\label{T_B1_T}\
\frac{{d\overline {{\xi ^*}} }}{{d{P_J}}} = \frac{{d\int_{\rm{0}}^\infty  {{\xi ^*} \times \varphi\left( x \right)dx} }}{{d{P_J}}} = \int_{\rm{0}}^\infty  {\frac{{d{\xi ^*}}}{{d{P_J}}} \times \varphi\left( x \right)} dx.
\end{align}
It is evident that the monotonic relationship between ${\overline {{\xi ^*}} }$ and $P_J$ depends on ${\frac{{d{\xi ^*}}}{{d{P_J}}}}$. First, we introduce auxiliary variables  $\theta  = \exp \left( { - \frac{{{P_J}\left( {1 - {\rho _{m,b}}} \right)}}{{{\rho _{a,b}}{P_a}\sigma _{A,B}^2}}x} \right)$, and  $\beta  =  - \left( {1 - \frac{{{\rho _{m,b}}{P_J}\sigma _{M,B}^2}}{{{\rho _{a,b}}{P_a}\sigma _{A,B}^2}}} \right)\theta  + 1$. Following that, ${\frac{{d{\xi ^*}}}{{d{P_J}}}}$ is given by
\begin{align}\label{T_B1_4T}\
\frac{{d{\xi ^*}}}{{d{P_J}}} = \exp \left( { - \frac{{{k_1}}}{{{\rho _{a,b}}{P_a}\sigma _{A,B}^2}}} \right)\left( { - {{\left( {\frac{{{\rho _{a,b}}{P_a}\sigma _{A,B}^2}}{{{\rho _{m,b}}{P_J}\sigma _{M,B}^2}} - 1} \right)}^{ - 2}}\frac{{{\rho _{a,b}}{P_a}\sigma _{A,B}^2}}{{{P_J}{\rho _{m,b}}{P_J}\sigma _{M,B}^2}}\left( {\beta \ln \beta  + 1 - \beta } \right) + \frac{{d\theta }}{{d{P_J}}}} \right).
\end{align}

It is evident that $\frac{{d\theta }}{{d{P_J}}} < 0$ and ${\beta \ln \beta  + 1 - \beta } \ge 0$. Thus, $\frac{{d{\xi ^*}}}{{d{P_J}}} < 0$. The proof is completed.

\section{}
\subsection{The monotonicity of $\overline {{\xi ^*}}$ with $\rho_{a,b}$}
Since ${{\rho _{a,b}}}$ has no effect on $\varphi\left( x \right)$, the derivative $\frac{{d \overline {{\xi ^*}} }}{{d {{\rho _{a,b}}}}}$ can be written as
\begin{align}\label{T_B1_of_ab}\
\frac{{d\overline {{\xi ^*}} }}{{d{\rho _{a,b}}}} = \frac{{d\int_{\rm{0}}^\infty  {{\xi ^*} \times \varphi \left( x \right)dx} }}{{d{\rho _{a,b}}}} = \int_{\rm{0}}^\infty  {\frac{{d{\xi ^*}}}{{d{\rho _{a,b}}}} \times \varphi \left( x \right)} dx.
\end{align}
By following similar procedures as in Appendix A, $\frac{{\partial {\xi ^*}}}{{\partial {\lambda _1}}}$ is given by
\begin{align}\label{T_B1_5T_ab}\
\frac{{d{\xi ^*}}}{{d{\rho _{a,b}}}} = \frac{1}{{{\rho _{a,b}}}}\exp \left( { - \frac{{{k_1}}}{{{\rho _{a,b}}{P_a}\sigma _{A,B}^2}}} \right)\left( {{{\left( {\frac{{{\rho _{m,b}}{P_J}\sigma _{M,B}^2}}{{{\rho _{a,b}}{P_a}\sigma _{A,B}^2}} - 1} \right)}^{ - 2}}\frac{{{\rho _{m,b}}{P_J}\sigma _{M,B}^2}}{{{\rho _{a,b}}{P_a}\sigma _{A,B}^2}}\left( {\beta \ln \beta  + 1 - \beta } \right) - \theta \ln \theta } \right).
\end{align}

It is obvious that $0 < \theta  < 1$, thus ${\theta \ln \theta }<0$, and ${\beta \ln \beta  + 1 - \beta } \ge 0$. Hence, we have $\frac{{d \overline {{\xi ^*}} }}{{d {\rho_{a,b}}}} > 0$. The proof is completed.
\subsection{The monotonicity of $\overline {{\xi ^*}}$ with $\rho_{m,b}$}
Since $\rho_{m,b}$ has no effect on $\varphi\left( x \right)$, the derivative $\frac{{d\overline {{\xi ^*}} }}{{d{\rho_{m,b}}}}$ can be written as
\begin{align}\label{T_B1_T}\
\frac{{d\overline {{\xi ^*}} }}{{d{\rho _{m,b}}}} = \frac{{d\int_{\rm{0}}^\infty  {{\xi ^*} \times \varphi \left( x \right)dx} }}{{d{\rho _{m,b}}}} = \int_{\rm{0}}^\infty  {\frac{{d{\xi ^*}}}{{d{\rho _{m,b}}}} \times \varphi \left( x \right)} dx.
\end{align}
Then, ${\frac{{d{\xi ^*}}}{{d{\rho_{m,b}}}}}$ is given by
\begin{align}\label{T_B1_5T}\
\frac{{d{\xi ^*}}}{{d{\rho _{m,b}}}} = \exp \left( { - \frac{{{k_1}}}{{{\rho _{a,b}}{P_a}\sigma _{A,B}^2}}} \right)\left( { - {{\left( {\frac{{{\rho _{a,b}}{P_a}\sigma _{A,B}^2}}{{{\rho _{m,b}}{P_J}\sigma _{M,B}^2}} - 1} \right)}^{ - 2}}\frac{{{\rho _{a,b}}{P_a}\sigma _{A,B}^2}}{{{\rho _{m,b}}{\rho _{m,b}}{P_J}\sigma _{M,B}^2}}\left( {\beta \ln \beta  + 1 - \beta } \right) + \frac{{d\theta }}{{d{\rho _{m,b}}}}} \right).
\end{align}

It is evident that $\frac{{d\theta }}{{d{\rho_{m,b}}}} >0$ and ${\beta \ln \beta  + 1 - \beta } \ge 0$. Thus,
there exists a value of $\rho_{m,b}$ that can achieve $\frac{{d{\xi ^*}}}{{d{\rho _{m,b}}}} >0$, and $\overline {{\xi ^*}} $ increases with $\rho_{m,b}$. Also. there exists a value of $\rho_{m,b}$ such that $\overline {{\xi ^*}} $ decreases with $\rho_{m,b}$. Hence, $\rho_{m,b}$ causes non-monotonic influence on $\overline {{\xi ^*}} $. The proof is completed.

\section{}

According to \eqref{ENP}, $E\left[ {\bf{X}} \right]$ can be derived as
\begin{align}\label{T_B1_T}\
E\left[ {\bf{X}} \right] = Prob\left( {\frac{{{P_a}{{\left| {{h_{A,M}}} \right|}^2}}}{{\eta {P_J}{{\left| {{h_{M,M}}} \right|}^2} + \sigma _m^2}} \ge \frac{{{P_a}{{\left| {{{\hat h}_{A,B}}} \right|}^2}}}{{{P_a}{{\left| {{{\widetilde h}_{A,B}}} \right|}^2} + {P_J}{{\left| {{h_{M,B}}} \right|}^2} + \sigma _b^2}}} \right).
\end{align}
Let  ${\gamma _M} = \frac{{{P_a}{{\left| {{h_{A,M}}} \right|}^2}}}{{\eta {P_J}{{\left| {{h_{M,M}}} \right|}^2} + \sigma _m^2}}$, and the CDF and PDF of ${{\gamma _M}}$  are, respectively, given by
\begin{align}\label{CDFgamma_m}\
{F_{{\gamma _M}}}\left( {{z}} \right)&= \int_0^\infty  {\frac{1}{{{P_J}\sigma _{M,M}^2}}\exp \left( { - \frac{x}{{{P_J}\sigma _{M,M}^2}}} \right)} dx\int_0^{{z}\left( {\eta x + \sigma _m^2} \right)} {\frac{1}{{{P_a}\sigma _{A,M}^2}}\exp \left( { - \frac{y}{{{P_a}\sigma _{A,M}^2}}} \right)} dy\nonumber \\
 &= 1 - \frac{1}{{{P_J}\sigma _{M,M}^2}}\frac{1}{{\left( {\frac{1}{{{P_J}\sigma _{M,M}^2}} + \frac{{{z}\eta }}{{{P_a}\sigma _{A,M}^2}}} \right)}}\exp \left( { - \frac{{{z}\sigma _m^2}}{{{P_a}\sigma _{A,M}^2}}} \right)
\end{align}
and
\begin{align}\label{PDFgamma_m}\
{f_{{{{\gamma _M}}}}}\left( {{z}} \right) &= \frac{1}{{{P_J}\sigma _{M,M}^2}}\frac{1}{{{P_a}\sigma _{A,M}^2}}\exp \left( { - \frac{{{z}\sigma _m^2}}{{{P_a}\sigma _{A,M}^2}}} \right){\left( {\frac{1}{{{P_J}\sigma _{M,M}^2}} + \frac{{{z}\eta }}{{{P_a}\sigma _{A,M}^2}}} \right)^{ - 1}}\nonumber \\
 &\times \left( {\eta {{\left( {\frac{1}{{{P_J}\sigma _{M,M}^2}} + \frac{{{z}\eta }}{{{P_a}\sigma _{A,M}^2}}} \right)}^{ - 1}} + \sigma _m^2} \right).
\end{align}

Similarly, let ${\gamma _B} = \frac{{{P_a}{{\left| {{{\hat h}_{A,B}}} \right|}^2}}}{{{P_a}{{\left| {{{\widetilde h}_{A,B}}} \right|}^2} + {P_J}{{\left| {{h_{M,B}}} \right|}^2} + \sigma _b^2}}$, and the CDF of ${{\gamma _B}}$ is given by
\begin{align}\label{CDFgamma_b}\
{F_{{\gamma _B}}}\left( z \right) &= 1 - \frac{{{{\left( {\frac{1}{{{\rho _{a,b}}{P_a}\sigma _{A,B}^2}} + \frac{z}{{\left( {1 - {\rho _{a,b}}} \right){P_a}\sigma _{A,B}^2}}} \right)}^{ - 1}} - {{\left( {\frac{1}{{{P_J}\sigma _{M,B}^2}} + \frac{z}{{\left( {1 - {\rho _{a,b}}} \right){P_a}\sigma _{A,B}^2}}} \right)}^{ - 1}}}}{{{\rho _{a,b}}{P_a}\sigma _{A,B}^2 - {P_J}\sigma _{M,B}^2}}\nonumber\\
 &\times \exp \left( { - \frac{{z\sigma _b^2}}{{\left( {1 - {\rho _{a,b}}} \right){P_a}\sigma _{A,B}^2}}} \right).
\end{align}
By invoking \cite[eq. (3.352.4)]{IS} and \cite[eq. (3.353.3)]{IS}, $E\left[ {\bf{X}} \right]$ can be calculated as
\begin{small}
\begin{align}\label{ENP_exact}\
E\left[ {\bf{X}} \right] &= Prob\left( {{\gamma _M} \ge {\gamma _B}} \right) = \int_0^\infty  {{F_{{\gamma _B}}}\left( x \right){f_{{\gamma _M}}}\left( x \right)} dx\nonumber\\
 &= 1 - \frac{1}{{{P_J}\sigma _{M,M}^2{P_a}\sigma _{A,M}^2}}\int_0^\infty  {\exp \left( { - \left( {\frac{{\sigma _b^2}}{{\left( {1 - {\rho _{a,b}}} \right){P_a}\sigma _{A,B}^2}} + \frac{{\sigma _m^2}}{{{P_a}\sigma _{A,M}^2}}} \right)x} \right){{\left( {\frac{1}{{{P_J}\sigma _{M,M}^2}} + \frac{{x\eta }}{{{P_a}\sigma _{A,M}^2}}} \right)}^{ - 1}}} \nonumber\\
 &\times \left( {\eta {{\left( {\frac{1}{{{P_J}\sigma _{M,M}^2}} + \frac{{x\eta }}{{{P_a}\sigma _{A,M}^2}}} \right)}^{ - 1}} + \sigma _m^2} \right){\left( {1 + \frac{{{\rho _{a,b}}}}{{1 - {\rho _{a,b}}}}x} \right)^{ - 1}}{\left( {1 + \frac{{{P_J}\sigma _{M,B}^2}}{{\left( {1 - {\rho _{a,b}}} \right){P_a}\sigma _{A,B}^2}}x} \right)^{ - 1}}dx\nonumber\\
 &= 1 - \left( {\frac{{\sigma _b^2}}{{\left( {1 - {\rho _{a,b}}} \right){P_a}\sigma _{A,B}^2}} - {{\left( {\frac{{{P_a}\sigma _{A,M}^2}}{{\eta {P_J}\sigma _{M,M}^2}} - \frac{{1 - {\rho _{a,b}}}}{{{\rho _{a,b}}}}} \right)}^{ - 1}} - {{\left( {\frac{{{P_a}\sigma _{A,M}^2}}{{\eta {P_J}\sigma _{M,M}^2}} - \frac{{\left( {1 - {\rho _{a,b}}} \right){P_a}\sigma _{A,B}^2}}{{{P_J}\sigma _{M,B}^2}}} \right)}^{ - 1}}} \right)\nonumber\\
 &\times {\left( {\frac{{\sigma _{A,M}^2\sigma _{M,B}^2}}{{\eta \sigma _{M,M}^2\left( {1 - {\rho _{a,b}}} \right)\sigma _{A,B}^2}} - 1} \right)^{ - 1}}{\left( {\frac{{{\rho _{a,b}}}}{{\left( {1 - {\rho _{a,b}}} \right)}} - \frac{{\eta {P_J}\sigma _{M,M}^2}}{{{P_a}\sigma _{A,M}^2}}} \right)^{ - 1}}L\left( {\frac{{\sigma _{A,M}^2\sigma _b^2}}{{\eta \left( {1 - {\rho _{a,b}}} \right){P_J}\sigma _{M,M}^2\sigma _{A,B}^2}} + \frac{{\sigma _m^2}}{{\eta {P_J}\sigma _{M,M}^2}}} \right)\nonumber\\
 &+ \frac{1}{{\left( {{P_a}\sigma _{A,B}^2{\rho _{a,b}} - {P_J}\sigma _{M,B}^2} \right)}}{\left( {\frac{{\sigma _{A,M}^2}}{{\left( {1 - {\rho _{a,b}}} \right)\sigma _{A,B}^2}} - \frac{{\eta {P_J}\sigma _{M,M}^2}}{{{\rho _{ab}}{P_a}\sigma _{A,B}^2}}} \right)^{ - 1}}\nonumber\\
 &\times \left( {{{\left( {\frac{1}{{\eta {P_J}\sigma _{M,M}^2}} - \frac{{1 - {\rho _{a,b}}}}{{{\rho _{a,b}}{P_a}\sigma _{A,M}^2}}} \right)}^{ - 1}} + \sigma _m^2} \right)L\left( {\frac{{\sigma _b^2}}{{{\rho _{a,b}}{P_a}\sigma _{A,B}^2}} + \frac{{\left( {1 - {\rho _{a,b}}} \right)\sigma _m^2}}{{{\rho _{a,b}}{P_a}\sigma _{A,M}^2}}} \right)\nonumber\\
 &- \frac{1}{{\left( {{P_a}\sigma _{A,B}^2{\rho _{a,b}} - {P_J}\sigma _{M,B}^2} \right)}}{\left( {\frac{{\sigma _{A,M}^2}}{{\left( {1 - {\rho _{a,b}}} \right)\sigma _{A,B}^2}} - \frac{{\eta \sigma _{M,M}^2}}{{\sigma _{M,B}^2}}} \right)^{ - 1}}\nonumber\\
 &\times \left( {{{\left( {\frac{1}{{\eta {P_J}\sigma _{M,M}^2}} - \frac{{\left( {1 - {\rho _{a,b}}} \right)\sigma _{A,B}^2}}{{{P_J}\sigma _{M,B}^2\sigma _{A,M}^2}}} \right)}^{ - 1}} + \sigma _m^2} \right)L\left( {\frac{{\sigma _b^2}}{{{P_J}\sigma _{M,B}^2}} + \frac{{\left( {1 - {\rho _{a,b}}} \right)\sigma _{A,B}^2\sigma _m^2}}{{{P_J}\sigma _{M,B}^2\sigma _{A,M}^2}}} \right)\nonumber\\
 &- {\left( {\frac{{{P_a}\sigma _{A,M}^2\sigma _{M,B}^2}}{{\eta \left( {1 - {\rho _{a,b}}} \right){P_a}\sigma _{M,M}^2\sigma _{A,B}^2}} - 1} \right)^{ - 1}}{\left( {\frac{{{\rho _{a,b}}{P_a}\sigma _{A,M}^2}}{{\eta \left( {1 - {\rho _{a,b}}} \right){P_J}\sigma _{M,M}^2}} - 1} \right)^{ - 1}},
\end{align}
\end{small}
where $L\left( x \right) = \exp \left( x \right)Ei\left( { - x} \right)$.

\end{appendices}

\bibliography{covert}

\end{document}